\documentclass[twocolumn,times]{aastex631}
\usepackage{xspace}
\usepackage{soul}

\newcommand{\msun}{$M_{\odot}$}
\newcommand{\teff}{${T}_{\mathrm{eff}}$}

\usepackage{array}
\newcolumntype{H}{>{\setbox0=\hbox\bgroup}c<{\egroup}@{}}

\shorttitle{Variable White Dwarfs in Gaia DR3}
\shortauthors{Steen et al.}
\begin{document}

\title{Measuring White Dwarf Variability from Sparsely Sampled Gaia DR3 Multi-Epoch Photometry}

\author[0009-0002-0435-5055]{Maya Steen}
\affiliation{Department of Astronomy, Boston University, 725 Commonwealth Ave., Boston, MA 02215, USA}
\affiliation{Department of Astronomy, New Mexico State University, Las Cruces, NM 88003, USA}

\author[0000-0001-5941-2286]{J. J. Hermes}
\affiliation{Department of Astronomy, Boston University, 725 Commonwealth Ave., Boston, MA 02215, USA}

\author[0000-0001-9632-7347]{Joseph A. Guidry}\altaffiliation{NSF Graduate Research Fellow}
\affiliation{Department of Astronomy, Boston University, 725 Commonwealth Ave., Boston, MA 02215, USA}

\author{Annabelle Paiva}
\affiliation{Department of Astronomy, Boston University, 725 Commonwealth Ave., Boston, MA 02215, USA}

\author[0000-0003-1748-602X]{Jay Farihi}
\affiliation{Department of Physics \& Astronomy, University College London, Gower Street, London WC1E 6BT, UK}

\author[0000-0003-3868-1123]{Tyler M. Heintz}
\affiliation{Department of Astronomy, Boston University, 725 Commonwealth Ave., Boston, MA 02215, USA}

\author[0009-0008-4954-8807]{Brison B. Ewing}
\affiliation{Department of Astronomy, Boston University, 725 Commonwealth Ave., Boston, MA 02215, USA}

\author[0000-0002-7461-8669]{Nathaniel Berry}
\affiliation{Department of Physics and Astronomy, University of North Carolina, Chapel Hill, NC 27599, USA}
 
\begin{abstract}

White dwarf stars are ubiquitous in the Galaxy, and are essential to understanding stellar evolution. While most white dwarfs are photometrically stable and reliable flux standards, some can be highly variable, which can reveal unique details about the endpoints of low-mass stellar evolution. In this study we characterize a sample of high-confidence white dwarfs with multi-epoch photometry from Gaia Data Release 3. We compare these Gaia light curves with light curves from the Zwicky Transiting Facility and the Transiting Exoplanet Survey Satellite to see when Gaia data independently can accurately measure periods of variability. From this sample, 105 objects have variability periods measured from the Gaia light curves independently, with periods as long as roughly 9.5\,d and as short as 256.2\,s (roughly 4\,min), including seven systems with periods shorter than 1000\,s. We discover 86 new objects from the 105 target sample, including pulsating, spotted, and binary white dwarfs, and even a new 68.4\,min eclipsing cataclysmic variable. The median amplitude of the absolute photometric variability we confirm from Gaia independently is 1.4\%, demonstrating that Gaia epoch photometry is capable of measuring short-term periods even when observations are sparse. 
\end{abstract}

\section{Introduction} \label{sec:intro}

\begin{deluxetable*}{ccc}
\tablecaption{Summary table of the full sample and a subset of Gaia-detected variable candidate white dwarfs.
\label{tab:summary}}
\tablewidth{20pt}
\tabletypesize{\small}
\tablehead{
\colhead{ } & \colhead{Full Sample} & \colhead{Gaia Period Match} }
\startdata
Variable objects near white dwarfs in CMD ("Variable candWDs") & 1333 & 105 \\
Candidate WDs with $P_{\rm WD} > 0.75$ & 1217 & 104 \\
Variable candWDs with spectroscopic confirmation & 568 & 36 \\
Variable candWDs that are known binaries & 91 & 8 \\
Variable candWDs that are known pulsating WDs & 49 & 2 \\
Variable candWDs known to have magnetic spots & 12 & 8 \\
Variable candWDs with TESS data & 698 & 65 \\
Variable candWDs with ZTF data & 992 & 72 \\
\enddata
\end{deluxetable*}

As the endpoints of all stars below roughly 8\,\msun, white dwarfs are extremely common. Their ubiquity and relative simplicity mean that white dwarf stars play an important role as flux standards, which are needed for precise absolute photometry and spectroscopy \citep[e.g.,][]{2022ApJ...940...19C}.

Roughly 97\% of all apparently isolated white dwarfs are empirically stable to less than 1\% variability on weeks-long timescales \citep{2017MNRAS.468.1946H}. When white dwarfs are photometrically variable they can reveal unique physical properties such as the presence of magnetism or surface rotation periods  \citep[e.g.,][]{2015MNRAS.447.1749M}, binary properties \citep[e.g.,][]{2011AJ....142...62H}, remnant planetary systems \citep[e.g,][]{2018MNRAS.476..933H,2021ApJ...912..125G}, or even their interiors through asteroseismology \citep[e.g.,][]{2012MNRAS.420.1462R}. 

Binary systems with at least one white dwarf star can tell us about stellar ages and close binary evolution (e.g., \citealt{2007MNRAS.382.1377R,2022ApJ...934..148H}). These objects can also be progenitors of Type Ia supernovae, which are important distance indicators \citep{2015MNRAS.447.1749M}. Binaries with short periods (less than roughly 1 hour) are also potential candidates for gravitational waves detectable with the Laser Interferometer Space Antenna (LISA, \citealt{2020ApJ...905...32B}). 

There are a number of ways binaries with a white dwarf can be photometrically variable. Eclipsing systems are the most dramatic, where one star passes in front of the other (e.g., \citealt{2016MNRAS.458..845H}). Eclipsing white dwarfs, which can be used to measure the masses and radii of the system components \citep{2017MNRAS.470.4473P,2018MNRAS.481.1083P}, are still relatively rare so any increase in the sample is valuable. Eclipses are not necessary to detect close binaries, as hot stars in the system can create a reflection effect from reprocessed heating off a cool companion (e.g., \citealt{2022A&A...666A.182S}). In addition, cataclysmic variables (CVs) are highly variable white dwarfs with stochastic mass-transfer from a Roche-lobe-filling main-sequence companion \citep{1995cvs..book.....W}.


Magnetic white dwarfs can also exhibit strong photometric changes, such as cold (e.g., \citealt{2013ApJ...773...47B,2015ApJ...814L..31K}) or hot (e.g., \citealt{2018AJ....156..119H}) spots corresponding to the surface rotation period. Measuring these rotation periods can help to investigate the angular momentum loss during white dwarf formation \citep{2019MNRAS.485.3661F}.

Pulsating white dwarfs are variable due to non-radial pulsations as they cool through different instability strips that depend on their surface composition \citep{2008ARA&A..46..157W,2008PASP..120.1043F,2017ApJS..232...23H}. With enough identified pulsation modes, asteroseismology affords the opportunity to significantly constrain the interiors of white dwarfs, including global rotation rates (e.g., \citealt{2015MNRAS.451.1701H,2017ApJ...835..277H}) and core composition (e.g., \citealt{2018Natur.554...73G,2022ApJ...935...21C}).
 
Motivated by the need to find and constrain new variable white dwarfs, we analyze here a new data product from Gaia Data Release 3 (DR3), which features multi-epoch light curves for sources classified as variable via a machine learning algorithm trained on constant objects \citep{2023A&A...674A..13E}. This new Gaia DR3 dataset includes photometric time series in the Gaia $G$, $G_{\mathrm{BP}}$, and  $G_{\mathrm{RP}}$ bands. Previous efforts have focused on literature cross-matches of DR3 variables \citep{2023A&A...674A..22G,2023A&A...674A..14R}, as well as specific object types \citep{2023A&A...674A..36G,2023A&A...674A..21M}.

In Section~\ref{sec:obs} we discuss our target selection for variable white dwarfs in Gaia DR3, and in Section~\ref{sec:analysis} describe our period-finding analysis. In Section~\ref{sec:wdvariables} we explain our classification of new white dwarf variables, including new (eclipsing) binaries, spotted (magnetic) white dwarfs, and pulsating white dwarfs, all with coherent periodicities detected from Gaia DR3 photometry independently and confirmed in multiple datasets. We discuss the strengths and limitations of Gaia for period analysis in Section~\ref{sec:gaiaperiods} and conclude in Section~\ref{sec:summary}.

\begin{figure*}[t]
\centering{\includegraphics[width=0.995\textwidth]{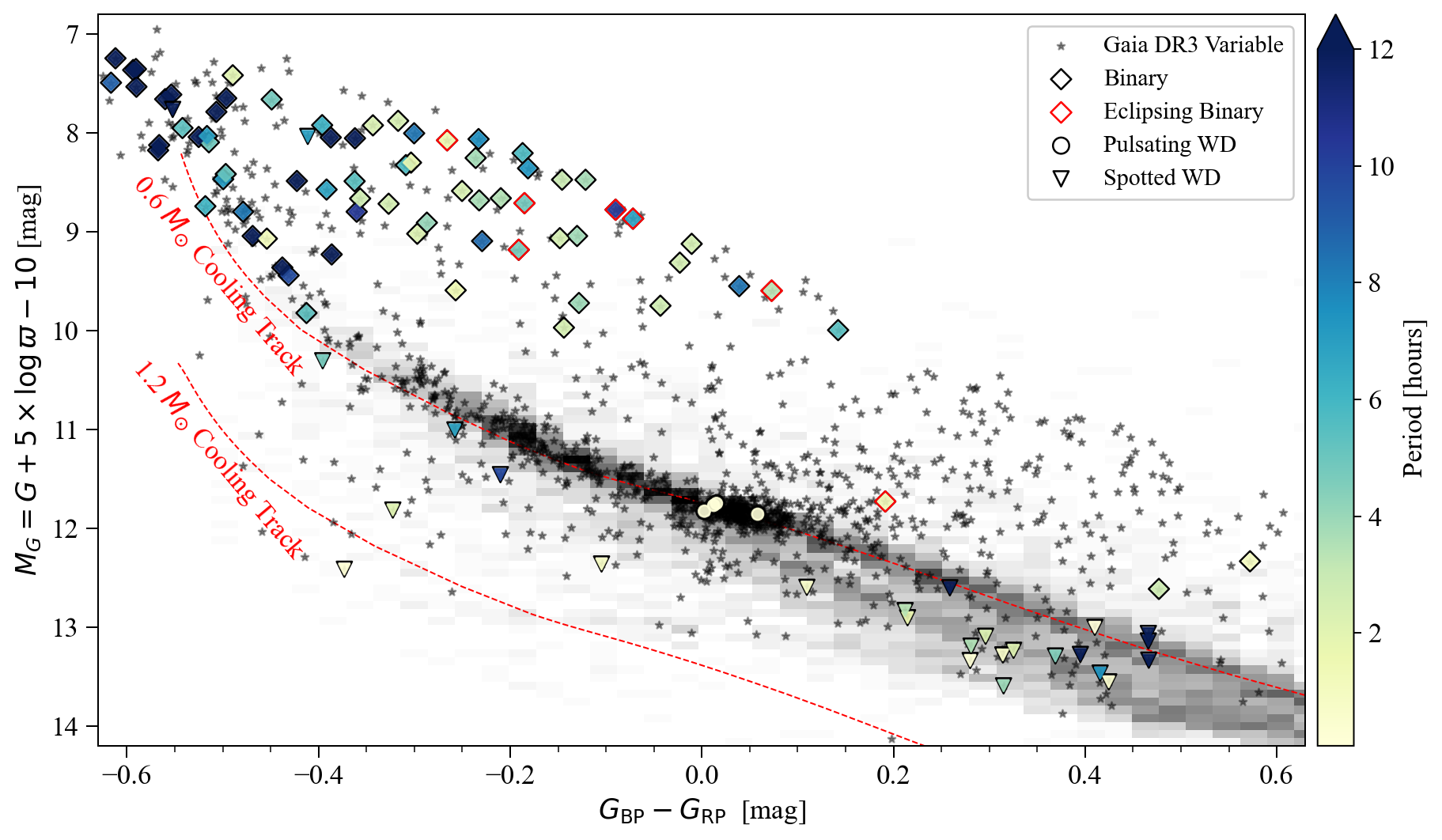}}
\caption{Color-magnitude diagram showing our sample of 1333 high-confidence white dwarf candidates flagged as photometrically variable in Gaia DR3 as black stars. The larger colored symbols show the 105 objects discussed in this manuscript which have variability periods measured from their Gaia light curves independently; the color for each point corresponds to the corresponding variability period. The background of this plot is a two-dimensional histogram of white dwarfs within 200 pc matching the magnitude distribution of the 1333 photometrically variable white dwarfs.  Dashed red lines show the expected cooling tracks of a 0.6\,\msun\ white dwarf at top and 1.2\,\msun\ white dwarf at bottom; white dwarfs cool from the upper left to the lower right in this diagram.  \label{fig:cmd}}
\end{figure*} 

\section{Target Selection and Observations} \label{sec:obs}

We set out to select and analyze all apparently isolated white dwarf stars flagged as photometrically variable in Gaia DR3 \citep{2023A&A...674A...1G}. A new, expanded data product in DR3 is the release of multi-epoch, multi-color ($G$, $G_{\mathrm{BP}}$, $G_{\mathrm{RP}}$) light curves for roughly 10.5 million objects with anomalously large scatter in their many Gaia visits (a.k.a. transits) that make them high-probability candidate variables \citep{2023A&A...674A..13E}. These objects are cataloged in Gaia DR3 with the entry {\tt phot\_variable\_flag} set to {\tt VARIABLE}; full light curves are available through the Gaia Datalink Service \citep{cd2022}, which we accessed using {\tt astroquery.gaia} \citep{2019AJ....157...98G}. 

In order to focus on isolated white dwarfs, we first restrict the 10.5 million variables in DR3 to be near the white dwarf region of the Gaia color-magnitude diagram by requiring $M_G > (4.5 \times G_{\mathrm{BP}}-G_{\mathrm{RP}}) + 9.0$, similar to the absolute magnitude selection of \citet{2018MNRAS.480.3942H}. This excludes many white dwarfs in close binaries, such as CVs \citep{2020MNRAS.494.3799P} and post-common-envelope (WD+dM) binaries.

We then cross-match this down-selected sample with the Gaia DR3 catalog of candidate white dwarfs of \citet{2021MNRAS.508.3877G}. This cross-match results in a final sample of 1333 objects, 1217 of which are high-confidence single white dwarfs (with $P_{{\mathrm WD}} > 0.75$, see Table \ref{tab:summary}); only 42 objects have $P_{{\mathrm WD}} < 0.3$.  We expect minor contamination from our color cuts from non-isolated white dwarfs, especially CVs with low-luminosity donors.

We further cross-match this sample with the Montreal White Dwarf Database \citep{2017ASPC..509....3D} and SIMBAD \citep{2000A&AS..143....9W}  by matching coordinates for each source (precessing the Gaia DR3 epoch to J2000). This cross-match was done in order to collect as much supporting information as possible including literature references and spectroscopic classifications. Relevant information found in this cross match is listed in Table~\ref{tab:2}.

We compare the Gaia multi-color light curves of our 1333 objects with two other long-baseline photometric surveys. First, we use the Python package {\tt Lightkurve} \citep{2018ascl.soft12013L,saunders_2020} to download any 20-second- and 2-min-cadence PDCSAP light curves of our objects as seen by the Transiting Exoplanet Survey Satellite (TESS, \citealt{2015JATIS...1a4003R}). This provides at least one month of nearly continuous TESS data for 698 objects in our sample, most brighter than roughly $G=17.0$\,mag.

Where available, we also download light curves within a 1.4-arcsecond search radius of our targets from Data Release 12 of the Zwicky Transient Facility (ZTF, \citealt{2019PASP..131a8002B}). This provides long-baseline ZTF observations for 992 objects across all magnitudes in our sample, all with declinations north of $-20$\,degrees. We also search for light curves in the south from the All-Sky Automated Survey for SuperNovae (ASAS-SN, \citealt{2014ApJ...788...48S,2017PASP..129j4502K}) and the Asteroid Terrestrial-impact Last Alert System (ATLAS, \citealt{2018PASP..130f4505T}) projects. However, any objects without TESS photometry in the south mostly had upper limits on photometry due to their faintness, so we did not use their light curves in our analysis here. 

\section{Period-Finding Analysis} \label{sec:analysis}

We analyze the Gaia light curves and ZTF and TESS light curves, when available, simultaneously. All available light curves are visually inspected for signatures of variability in both the time and frequency domain, informed by the spectroscopic classification of the white dwarf if known.

We analyze every light curve in the frequency domain by computing Lomb-Scargle periodograms \citep{2013A&A...558A..33A} using {\tt astropy} \citep{2013A&A...558A..33A,2018AJ....156..123A, 2022ApJ...935..167A}. We also separate the light curves by band-pass filters ($G$, $G_{\mathrm{BP}}$, $G_{\mathrm{RP}}$, $g$, $r$, $T$ from Gaia, ZTF, and TESS, respectively) to search for color-dependant variability. We initially classify frequencies as significant if their amplitude exceeds five times the mean amplitude of all frequencies in the periodogram, a relatively conservative limit given prior sporadically sampled datasets \citep{1993A&A...271..482B,2021ApJ...912..125G}. For simplicity, we use a constant frequency step size for Gaia and ZTF data: $10^{-8}$\,Hz, which corresponds to a critical sampling of 1157 days, consistent with the length of the typical Gaia and ZTF datasets. The step size for the TESS periodogram frequencies is set by the inverse of the TESS light curve baseline, oversampled by a factor of 10. All periodograms are computed to the TESS 2-min Nyquist frequency of 4160\,$\mu$Hz.

We use all available survey information in addition to our coherent period analysis to attempt to classify as many variable candidate white dwarfs in our sample as possible (see Figure~\ref{fig:cmd} and top-line summary in Table~\ref{tab:summary}). The amount of information differs significantly from source to source. For example, only 586 of 1333 objects have a spectral type in the literature listed from either MWDD \citep{2017ASPC..509....3D} or SIMBAD \citep{2000A&AS..143....9W}. Therefore, our classification for 86 of 105 objects in Table~\ref{tab:2} listed as ``likely" is based on the limited information available and requires spectroscopic follow-up to truly confirm the variability classification.

\begin{longrotatetable} 
\movetabledown=16mm
\begin{deluxetable*}{cccccccccccccccccccccc}
\tablecaption{Literature information for the 105 white dwarfs in this study with significant variability measured from their Gaia DR3 light curve.}
\label{tab:2}
\tabletypesize{\tiny}
\tablehead{
\colhead{Class}  & \colhead{Gaia Source ID} & \colhead{WD J Name$^{\dagger}$} & \colhead{RA} & \colhead{DEC} &  \colhead{$G$} &  \colhead{$P_{\rm WD}$$^{\dagger}$} & \colhead{\teff$^{\dagger}$} & \colhead{Mass$^{\dagger}$}  & \colhead{Sp. Type}  & \colhead{Dist.} & \colhead{Class$^{\ddagger}$} & \colhead{Score$^{\ddagger}$} & $N_{\rm G}$$^{\ddagger}$ & $N_{\rm g}$$^{\ast \ast}$ & Skew$^{\ddagger}$ & IQR$^{\ddagger}$ & $N_{\rm G,S}$$^{\ast}$ & Abbe$^{\ast}$ & $P_{\rm G,S}$$^{\ast}$  & \colhead{$P_{\rm lit}$} & \colhead{Ref.} \\
\colhead{(This Work)} & \colhead{(DR3)} &  & \colhead{(J2016)} & \colhead{(J2016)} & \colhead{[mag]} &  &  \colhead{[K]}  & \colhead{[\msun]} &  & \colhead{[pc]} & & &  &  & ($G$) & ($G$) &  &   & \colhead{[hr]} & \colhead{[hr]} & 
} 
\startdata 
Likely Binary&2780903539324180736&WD J003213.13+160434.78&8.054739&16.076395&15.7&0.982&&&DO&415.3&WD&0.6&50&68.0&0.755&0.036&50.0&0.836&21.7869583&&1 \\ Likely Binary&2317319612801004416&WD J003449.86-312952.69&8.707866&-31.497956&16.1&0.882&25410.0&0.24&DA&431.1&&&39&&-0.416&0.133&39.0&0.822&2.3167499&&2 \\ Likely Binary&522432163970498048&WD J011053.06+604831.03&17.721109&60.808565&17.0&0.949&14770.0&0.2&&301.7&&&58&491.0&-0.088&0.068&58.0&0.872&8.3619614&& \\ Likely Binary&323342459647016448&WD J012828.99+385436.63&22.12078&38.910093&15.9&0.998&38310.0&0.54&DA&236.1&WD&0.651&30&375.0&-0.396&0.025&&&&&3 \\ Likely Binary&2478893842934875136&WD J013148.02-055900.14&22.950116&-5.983399&17.3&0.939&96630.0&0.63&&930.9&WD&0.465&44&329.0&-0.49&0.041&&&&& \\ Likely Binary&459237630076876672&WD J022704.24+591502.04&36.768226&59.25064&15.9&0.988&7360.0&0.51&DA&37.7&WD&0.52&48&420.0&0.282&0.029&&&&&4 \\ Likely Binary&4846154678323893248&WD J032205.10-472106.46&50.521313&-47.351787&17.3&0.978&&&&970.5&WD&0.501&48&&0.465&0.121&&&&& \\ Likely Binary&117770030481927936&WD J032803.02+265151.65&52.012578&26.864324&18.2&0.98&&&&656.2&WD&1.0&27&426.0&-0.095&0.16&&&&& \\ Likely Binary&235184643828418688&WD J032902.63+390340.87&52.260969&39.061346&18.6&0.973&&&&876.2&WD&0.5&41&251.0&-0.841&0.082&42.0&0.984&4.0387314&& \\ Likely Binary&5090605830056251776&WD J035814.52-213813.74&59.560498&-21.637194&17.8&0.938&24810.0&0.24&&802.3&&&76&242.0&-0.302&0.147&77.0&0.924&1.7951253&& \\ Likely Binary&4681729620697295360&WD J041033.37-585203.63&62.639044&-58.867634&17.6&0.94&15430.0&0.13&&659.0&&&39&&-0.386&0.175&39.0&0.902&2.8736527&& \\ Likely Binary&3204352783173513728&WD J043103.75-041702.63&67.765647&-4.284043&17.9&0.854&28220.0&0.24&&1021.7&&&34&68.0&3.032&0.171&34.0&0.911&2.432582&& \\ Likely Binary&3230486971974872192&WD J043832.74+003117.01&69.636517&0.521344&17.5&0.981&94280.0&0.75&&635.9&WD&0.437&53&570.0&-0.394&0.037&&&&& \\ Likely Binary&2974680502235811712&WD J050433.01-213007.93&76.137575&-21.502202&18.1&0.959&&&&1208.7&WD&0.252&51&171.0&-0.096&0.107&53.0&0.941&15.0267735&& \\ Likely Binary&3341182435406679424&WD J053632.07+123128.65&84.133666&12.524682&16.4&0.979&19510.0&0.25&&291.4&&&48&317.0&-0.347&0.141&50.0&0.906&3.8008657&& \\ Likely Binary&2941800362921100928&WD J061234.26-191111.75&93.142763&-19.18675&18.0&0.984&23960.0&0.33&&640.1&WD&0.5&48&188.0&-0.096&0.115&&&&2.2251&5 \\ Likely Binary&5549322119820590208&WD J061815.27-512305.52&94.563662&-51.384787&15.9&0.998&30420.0&0.5&&194.3&WD&0.639&36&&1.031&0.024&&&&& \\ Likely Binary&3125897813875563904&WD J064316.01+013012.78&100.816765&1.503518&15.8&0.996&93720.0&0.83&DO&251.9&WD&0.318&24&300.0&-0.773&0.06&24.0&0.817&6.0384346&&6 \\ Likely Binary&5575611644702244992&WD J064753.57-403756.16&101.973208&-40.632249&17.7&0.987&&&&862.7&&&43&&-0.232&0.15&44.0&0.831&7.0115164&& \\ Likely Binary&5286227228018008576&WD J064802.27-634629.90&102.009505&-63.774891&17.5&0.917&22690.0&0.23&&691.9&&&45&&-0.084&0.136&45.0&0.917&3.7701112&& \\ Likely Binary&1099005032089977984&WD J070647.52+613350.31&106.698306&61.563912&13.5&0.984&&&&158.0&&&64&915.0&-0.414&0.02&64.0&0.837&8.9701518&& \\ Likely Binary&989059981050056320&WD J072146.74+564408.41&110.444764&56.735656&18.8&0.983&13400.0&0.23&&651.7&&&32&41.0&0.164&0.093&32.0&0.947&2.7708059&& \\ Likely Binary&1085903148454238208&WD J073121.39+595737.01&112.839236&59.960251&16.1&0.988&&&DO&424.5&WD&0.273&43&550.0&-0.417&0.088&44.0&0.937&5.3138963&&7 \\ Likely Binary&5615892360564187008&WD J074316.04-224656.86&115.816834&-22.782453&19.0&0.831&38940.0&0.39&&1534.0&&&69&139.0&-0.59&0.196&71.0&0.919&8.289411&& \\ Likely Binary&5216924803961432192&WD J090240.78-742331.37&135.669175&-74.391845&17.2&0.881&10270.0&0.15&&278.8&WD&0.559&38&&-0.111&0.036&&&&& \\ Likely Binary&3842377248804727168&WD J091748.20+001041.72&139.450873&0.178195&17.4&0.998&34570.0&0.56&&406.2&WD&0.606&36&237.0&0.155&0.069&&&&& \\ Likely Binary&5437786045395659520&WD J093611.12-340059.16&144.046282&-34.016386&17.1&0.929&49990.0&0.41&&649.3&&&96&&-0.357&0.042&97.0&0.92&20.8380364&& \\ Likely Binary&3834895969825628800&WD J100645.76+003204.48&151.690524&0.534619&16.7&0.98&15890.0&0.21&DA&336.9&CV&0.382&26&238.0&-0.622&0.083&&&&&8 \\ Likely Binary&5456112705203412352&WD J105228.94-295308.20&163.120558&-29.885628&17.3&0.996&92120.0&0.83&&510.5&&&49&&-0.548&0.049&49.0&0.923&8.2170957&& \\ Likely Binary&5390573840730111872&WD J110318.00-401544.50&165.825071&-40.262433&15.1&0.982&&&&310.3&&&42&&-0.099&0.182&42.0&0.932&0.8834878&& \\ Likely Binary&5383832838018956032&WD J112355.21-400614.63&170.979607&-40.103915&14.9&0.979&&&&319.2&&&50&&-0.584&0.045&50.0&0.847&16.9484344&& \\ Likely Binary&5378958286359002624&WD J113832.79-435817.73&174.636565&-43.971582&16.5&0.981&&&&611.8&WD&0.552&61&&-0.176&0.076&61.0&0.893&13.2968409&& \\ Likely Binary&6076202559948311040&WD J121622.61-545907.30&184.094154&-54.98536&18.7&0.956&17070.0&0.24&&748.2&&&66&&-0.378&0.406&67.0&0.888&2.5750134&& \\ Likely Binary&1726297924930902400&WD J123955.40+834707.51&189.979829&83.785334&19.8&0.708&7020.0&0.27&&333.9&CV&0.716&27&32.0&0.544&0.796&&&&1.91383&9 \\ Likely Binary&3509435507186636288&WD J125005.11-203632.29&192.521286&-20.609046&17.4&0.992&&&&699.1&WD&0.577&29&222.0&-0.571&0.129&&&&& \\ Likely Binary&3721943209023827456&WD J135523.92+085645.42&208.849658&8.945966&17.5&0.979&26600.0&0.32&DA&594.8&&&42&402.0&-0.37&0.054&42.0&0.905&2.7448209&2.74511&10 \\ Likely Binary&1671880139535534208&WD J140330.89+671255.16&210.878613&67.215384&18.7&0.953&16210.0&0.17&&1042.8&&&43&615.0&-0.334&0.1&43.0&0.906&3.447414&& \\ Likely Binary&1286055427676135808&WD J144007.60+315413.02&220.031656&31.903652&17.3&0.992&&&&670.6&&&96&692.0&-0.477&0.055&97.0&0.978&23.2727485&& \\ Likely Binary&5773385817714683776&WD J145623.30-793654.99&224.096814&-79.615325&17.5&0.926&&&&934.8&&&58&&-0.314&0.155&58.0&0.92&4.8197997&& \\ Likely Binary&5873013322220969984&WD J145817.91-651024.94&224.574476&-65.173657&17.7&0.957&38640.0&0.42&&635.7&&&88&&-0.317&0.074&89.0&0.898&3.679142&& \\ Likely Binary&1728880879608190592&WD J153028.82+874848.48&232.613053&87.813576&15.6&0.99&&&DA&324.5&WD&0.605&48&&-0.306&0.056&&&&&11 \\ Likely Binary&1272870835853944704&WD J154012.07+290828.96&235.050333&29.141305&18.9&0.945&22570.0&0.25&DAH&1132.5&&&74&446.0&-0.019&0.226&74.0&0.848&2.3768165&&12 \\ Likely Binary&2121629147470776704&WD J181824.35+465721.46&274.601506&46.95593&17.5&0.937&18100.0&0.18&&640.1&&&37&1008.0&-0.253&0.175&37.0&0.903&3.7342113&7.46843&13 \\ Likely Binary&2095490942878080256&WD J183243.45+354518.42&278.181011&35.755123&18.1&0.994&24510.0&0.42&&503.6&WD&0.264&45&1602.0&-0.302&0.069&&&&& \\ Likely Binary&2144580907240525568&WD J184506.31+505412.15&281.276271&50.903336&17.8&0.911&18230.0&0.16&&830.7&&&45&1033.0&-0.512&0.105&45.0&0.955&5.7531055&& \\ Likely Binary&2126940892444693760&WD J191931.09+443243.88&289.879471&44.54546&16.6&0.926&33990.0&0.3&CV&535.7&WD&0.5&41&956.0&0.19&0.071&41.0&0.794&6.2207431&&14 \\ Likely Binary&1807309147099442432&WD J195952.94+143534.17&299.970566&14.592818&17.5&0.994&&&&643.8&&&51&329.0&-0.138&0.05&52.0&0.905&7.1919594&& \\ Likely Binary&1809460380271929344&WD J200937.02+165219.12&302.404376&16.872102&16.8&0.949&13230.0&0.14&&341.6&&&61&87.0&-0.119&0.166&62.0&0.938&2.49783&& \\ Likely Binary&2061416695278519808&WD J202034.57+392212.12&305.143844&39.369926&17.4&0.999&36520.0&0.72&&326.4&&&46&&-0.63&0.079&47.0&0.886&5.3152144&& \\ Likely Binary&6679158828046846336&WD J202354.90-433513.67&305.978719&-43.58712&17.0&0.976&&&&902.0&WD&0.515&49&&-0.07&0.107&&&&& \\ Likely Binary&1863757043284422784&WD J202841.38+325837.35&307.172333&32.976997&17.5&0.92&&&&758.4&&&56&453.0&-0.135&0.378&56.0&0.883&19.4392732&& \\ Likely Binary&6910676050839449728&WD J210110.17-052751.14&315.29212&-5.464319&14.8&0.998&36900.0&0.52&DBA&140.5&WD&0.662&58&37.0&0.21&0.025&58.0&0.868&12.8962353&&15 \\ Likely Binary&2177297391130977792&WD J210840.53+554502.63&317.168856&55.750555&18.9&0.939&8530.0&0.52&&180.9&&&53&613.0&-0.135&0.133&53.0&0.873&2.7846556&& \\ Likely Binary&1867851869401000960&WD J211557.58+364255.40&318.989974&36.715394&18.0&0.978&&&&773.2&&&60&477.0&-0.114&0.094&61.0&0.889&6.7479039&& \\ Likely Binary&1950573244454108160&WD J213119.93+345246.84&322.833067&34.879639&17.5&0.984&48120.0&0.54&&547.8&&&68&373.0&-0.601&0.073&69.0&0.985&10.3988838&& \\ Likely Binary&2667317799127474944&WD J213958.29-074126.17&324.992963&-7.690733&16.7&0.99&&&&517.2&WD&0.5&46&224.0&-0.197&0.093&&&&9.91934&16 \\ Likely Binary&2000785539622175104&WD J221545.15+503759.46&333.938098&50.633181&18.1&0.967&&&&847.4&&&54&55.0&0.126&0.129&54.0&0.913&5.4185123&& \\ Likely Binary&2833849800205759360&WD J224539.30+201633.35&341.414024&20.275965&15.7&0.994&14490.0&0.26&&157.0&WD&0.296&36&771.0&0.024&0.03&&&&& \\ Likely Binary&2608280729159380992&WD J224653.73-094834.48&341.724052&-9.809701&16.4&0.982&&&DA&525.7&WD&0.62&23&82.0&-0.134&0.038&&&&&17 \\ Likely Binary&6609580078678720512&WD J225206.39-274012.56&343.026633&-27.670175&16.7&0.993&72550.0&0.65&&447.8&&&32&268.0&-0.316&0.137&32.0&0.835&5.1362607&& \\ Likely Binary&2411529617359368960&WD J230047.06-125054.98&345.196258&-12.848608&16.7&0.995&34300.0&0.51&DB&310.9&WD&0.686&26&244.0&-0.148&0.053&&&&&18 \\ Likely Binary&2210180554096762624&WD J232411.16+653053.05&351.04663&65.514762&16.1&0.932&32760.0&0.27&&356.7&WD&1.0&54&558.0&0.109&0.108&54.0&0.858&7.2978052&& \\ Likely Binary&6523373973310305664&WD J233923.79-485350.19&354.849218&-48.897269&17.0&0.978&24160.0&0.27&&454.1&S&0.157&78&&-0.36&0.095&78.0&0.914&2.530295&& \\ Likely Eclipsing Binary&414813439007636352&WD J004502.35+503408.46&11.25983&50.568959&17.6&0.955&19200.0&0.23&&557.1&&&53&585.0&4.153&0.098&55.0&0.912&1.8351065&6.87792&19 \\ Likely Eclipsing Binary&6086116512688802176&WD J130905.43-470852.39&197.272675&-47.147918&18.3&0.908&11040.0&0.13&&542.4&&&49&&3.648&0.18&50.0&0.817&0.5338794&& \\ Likely Eclipsing Binary&6008982469163902464&WD J154008.28-392917.61&235.034874&-39.488217&17.3&0.973&9590.0&0.41&&131.9&CV&0.067&48&&4.052&0.095&49.0&0.738&0.6007567&& \\ Likely Eclipsing Binary&2056368017063855360&WD J202811.64+350525.08&307.048487&35.090314&17.9&0.955&20320.0&0.23&&670.1&&&52&610.0&-0.316&0.079&52.0&0.901&10.3938399&& \\ Known Eclipsing Binary&380560941677424768&WD J003352.64+385529.70&8.469299&38.924891&18.3&0.956&20640.0&0.24&&841.3&&&40&828.0&-0.285&0.113&40.0&1.002&4.8646611&4.86&20 \\ Known Eclipsing Binary&578539413395848704&WD J085746.18+034255.35&134.442416&3.715389&18.3&0.872&21040.0&0.19&DA&1088.4&&&72&209.0&1.739&0.235&73.0&0.874&1.5623108&1.562316667&21,22 \\ Known Eclipsing Binary&1446230735421615872&WD J132518.18+233808.02&201.325817&23.635475&18.2&0.983&17000.0&0.24&DA+dM&621.7&&&52&439.0&-0.328&0.077&52.0&0.944&4.6790375&4.679015783&23,24 \\ Known Binary&1112171030998592256&WD J071709.79+740040.53&109.292486&74.01089&15.0&0.996&14340.0&0.29&DA&101.3&WD&0.469&58&567.0&-0.331&0.025&58.0&0.796&2.4735485&2.473008&25 \\ Known Binary&5660008787157868928&WD J095017.97-251124.79&147.574789&-25.190355&15.5&0.888&19340.0&0.16&DA+dM&309.8&WD&1.0&42&&0.346&0.068&42.0&0.96&7.6478424&7.647696&26,27 \\ Known Binary&1223988576107848832&WD J153938.12+270605.84&234.908707&27.10159&16.6&0.955&24650.0&0.24&DA&447.5&WD&0.5&113&204.0&-0.027&0.044&116.0&0.865&5.7252728&5.725033333&28,29 \\ Likely CV&6180386608127234432&WD J130017.86-325805.23&195.074344&-32.968155&20.3&0.851&&&&590.8&CV&0.906&31&&-2.053&0.176&&&&&30 \\ Known CV&5099482805904892288&WD J031413.25-223542.92&48.555884&-22.595499&18.2&0.996&10710.0&0.66&CV&160.4&WD&0.578&31&79.0&-0.763&0.107&31.0&0.877&1.3503827&1.35&31,32 \\ Known CV&5645504991845223936&WD J083843.35-282701.14&129.680548&-28.450262&17.6&0.978&9500.0&0.41&CV&156.6&S&0.6&77&&0.232&0.847&77.0&0.513&16.1135388&1.640216667&33,34 \\ Likely DAV&5709854047291834496&WD J082924.77-163337.25&127.353247&-16.560327&17.6&0.996&10930.0&0.58&&141.4&&&40&140.0&-0.599&0.069&41.0&0.724&1.0316165&&35 \\ Likely DAV&3459790225027896448&WD J120650.89-380549.23&181.711597&-38.097216&15.7&0.998&11870.0&0.65&&58.8&WD&0.524&62&&-1.156&0.034&62.0&0.845&7.0349461&& \\ Known DAV&2026653131220381824&WD J195227.88+250929.18&298.116179&25.158763&15.1&0.998&11710.0&0.6&DA&47.4&&&79&393.0&-1.407&0.026&79.0&0.725&0.566988&0.071166667&36,37 \\ Known DAV&6410501923531836928&WD J215823.88-585353.81&329.599553&-58.898156&15.8&0.998&11640.0&0.61&DA&63.7&WD&0.761&62&&0.009&0.026&&&&0.086186111&38,39 \\ Likely Spot&313206676129937408&WD J011157.12+313632.67&17.987841&31.608892&16.9&0.999&18120.0&0.8&&125.3&WD&0.355&32&281.0&0.158&0.033&&&&& \\ Likely Spot&5139880551029408768&WD J013832.02-195446.58&24.634396&-19.912219&15.5&0.994&8430.0&0.93&DA&24.4&&&62&179.0&0.201&0.032&62.0&0.933&3.9428678&&40 \\ Likely Spot&105601048101988992&WD J020322.94+255851.90&30.845585&25.981003&18.1&0.992&9210.0&0.64&&124.9&WD&0.569&26&&0.695&0.129&&&&& \\ Likely Spot&4618860233015905408&WD J024235.52-820636.32&40.651134&-82.110184&16.4&0.995&9450.0&0.79&DA&53.1&WD&0.549&42&&0.711&0.035&42.0&0.91&3.3447333&&41 \\ Likely Spot&187576000803144704&WD J052026.35+381240.87&80.109816&38.211059&15.4&1.0&19830.0&0.69&DAB&74.6&WD&0.562&28&407.0&-0.391&0.017&&&&&42 \\ Likely Spot&2951380789991336192&WD J062555.04-141442.31&96.479499&-14.245324&16.5&0.994&8020.0&0.74&&43.9&WD&0.509&42&219.0&0.293&0.069&42.0&0.897&4.4630736&& \\ Likely Spot&3161618477052648192&WD J072758.87+101157.08&111.995427&10.198911&17.4&0.995&7600.0&0.77&&58.8&WD&0.331&24&286.0&0.242&0.027&&&&& \\ Likely Spot&1030289712881895296&WD J083531.17+533230.91&128.879468&53.541364&18.2&0.992&8490.0&0.8&DAH&96.7&WD&0.536&35&75.0&-0.382&0.065&&&&&43 \\ Likely Spot&5620763437599189376&WD J090212.89-394553.32&135.554166&-39.764881&16.0&0.993&8770.0&0.83&&36.4&WD&0.586&57&&-0.027&0.024&&&&& \\ Likely Spot&3598424931753893888&WD J115033.35-063618.07&177.637567&-6.605365&16.4&0.993&8770.0&0.89&DA&41.5&WD&0.509&48&130.0&0.259&0.026&&&&&44 \\ Likely Spot&1465214177336901376&WD J124740.10+303007.41&191.91707&30.501779&17.9&0.992&7460.0&0.65&DA&82.3&WD&0.632&41&548.0&-0.381&0.042&&&&&45 \\ Likely Spot&6279014454702485632&WD J144410.08-214338.33&221.041685&-21.727292&16.9&0.999&25120.0&1.12&&106.0&WD&0.777&22&102.0&0.454&0.045&&&&& \\ Likely Spot&1639907647071426176&WD J154801.37+623611.57&237.005479&62.603291&18.6&0.999&30150.0&0.75&&449.5&ECL&0.005&57&984.0&0.369&0.155&&&&9.474283333&46 \\ Likely Spot&4334641562477923712&WD J165335.21-100116.33&253.397409&-10.022141&15.7&0.991&7350.0&0.54&&32.6&WD&0.508&67&161.0&0.496&0.039&&&&& \\ Likely Spot&2043899619620928000&WD J185924.85+330157.53&284.853197&33.03259&17.7&0.993&9320.0&0.78&&91.8&WD&0.531&41&380.0&-0.083&0.089&&&&& \\ Likely Spot&2216598403463898112&WD J215432.26+634154.11&328.634861&63.698414&18.9&0.997&16590.0&1.03&&201.3&WD&0.4&38&542.0&-0.205&0.131&&&&& \\ Likely Spot&6613841171565998848&WD J221452.56-314333.98&333.719762&-31.726229&17.1&0.994&7670.0&0.75&&53.9&WD&0.515&42&&0.22&0.068&42.0&1.01&7.383116&& \\ Known Spot&2375576682347401216&WD J003512.88-122508.84&8.804346&-12.419793&16.9&0.993&8440.0&0.81&DAHe&52.7&WD&0.627&26&623.0&-0.107&0.041&&&&1.210483333&47 \\ Known Spot&4613612951211823104&WD J031715.85-853225.56&49.311809&-85.540486&14.8&1.0&26460.0&1.27&DAH&29.4&WD&0.614&46&&-0.146&0.136&46.0&0.607&0.6268161&0.201388889&48 \\ Known Spot&551153263105246208&WD J041246.85+754942.26&63.192633&75.828522&15.8&0.991&8600.0&0.76&DAHe&35.0&WD&0.525&51&418.0&0.249&0.037&52.0&0.933&2.2891284&2.2891&49,50 \\ Known Spot&504240778439455872&WD J053432.93+770757.40&83.637543&77.131959&16.5&0.995&10670.0&0.84&&60.7&&&48&410.0&0.176&0.074&48.0&0.909&0.7234655&0.72347&51 \\ Known Spot&1020641700210987520&WD J095125.94+530930.72&147.858294&53.158533&15.3&0.983&126070.0&0.78&DA&314.7&WD&0.596&54&397.0&-0.461&0.099&&&&82.8&52 \\ Known Spot&3682469122383597056&WD J125230.93-023417.72&193.129111&-2.571767&17.5&0.992&7740.0&0.56&DAHe&77.7&WD&0.555&58&&-0.362&0.054&&&&0.088132778&53 \\ Known Spot&4235280071072332672&WD J195629.23-010232.67&299.11983&-1.045547&13.6&0.994&7790.0&0.7&DA&11.6&WD&1.0&36&&0.206&0.059&&&&34.60319444&54,55 \\ Known Spot&1748833236082347136&WD J202932.51+070107.70&307.385483&7.018813&16.6&0.943&63100.0&0.52&DO&522.1&WD&0.294&45&183.0&-0.08&0.038&&&&6.96&56,57 \\ 
\enddata 
\tablerefs{$^{\dagger}$WD J Name, $P_{\rm WD}$ and atmospheric parameters determined from H-atmosphere fits of \citet{2021MNRAS.508.3877G}. Spectral type information collected from the Montreal White Dwarf Database and SIMBAD. All Gaia information arises from DR3 \citep{2023AA...674A...1G}. $^{\ddagger}$The best classification (Class) and classification score (Score) arise from the variability classification results of all classifiers table in DR3 ({\tt vclassre}). Summary statistics from the Gaia $G$ light curve such as the number of FoV transits used in analysis ($N_{\rm G}$), the skew, and the interquartile range (IQR) arise from the variability summary table in DR3 ({\tt varisum}).
$^{\ast \ast}$ Information from Zwickey Transient Facility, including the number of FoV transitys used in analysis from the $g$ band ($N_{\rm g}$) . $^{\ast}$Information for any objects flagged as short-timescale sources, such as the number of FoV transits used in the short-period analysis ($N_{\rm G,S}$), the mean of per-FoV Abbe values (Abbe), and the hypothesized timescale of variability ($P_{\rm G,S}$) arise from the short-timescale sources table in DR3 ({\tt vst}). Literature periods have been collected from (1) \citet{2016MNRAS.455.3413K}; (2) \citet{2001AA...378..556K}; (3) \citet{2013ApJS..204....5K}; (4) \citet{2020MNRAS.497..130T}; (5) \citet{2023ApJS..264...39R}; (6) \citet{1981AJ.....86.1384L}; (7) \citet{1995AA...303L..53D}; (8) \citet{2015MNRAS.446.4078K}; (9) \citet{2023ApJS..264...39R}; (10) \citet{2013MNRAS.429.3596B}; (11) \citet{2008AJ....136..946M}; (12) \citet{2015MNRAS.446.4078K}; (13) \citet{2023ApJS..264...39R}; (14) \citet{2013AJ....145..109H}; (15) \citet{2017MNRAS.472.4173R}; (16) \citet{2014ApJS..213....9D}; (17) \citet{2006ApJS..167...40E}; (18) \citet{2016MNRAS.459.4343K}; (19) \citet{2023ApJS..264...39R}; (20) \citet{2022MNRAS.516..720K}; (21) \citet{2016MNRAS.460.3873B}; (22) \citet{2015MNRAS.449.2194P}; (23) \citet{2010arXiv1009.3048D}; (24) \citet{2016MNRAS.460.3873B}; (25) \citet{1996MNRAS.278..565M}; (26) \citet{2017MNRAS.469.3688D}; (27) \citet{2020ApJ...894...53K}; (28) \citet{2016MNRAS.461.2747K}; (29) \citet{2010MNRAS.402..620R}; (30) \citet{10.1093/mnras/stu639}; (31) \citet{1979ApJ...234L.113B}; (32) \citet{1980MNRAS.191P..43W}; (33) \citet{2017ApJ...838..124H}; (34) \citet{2017MNRAS.471.2902R}; (35) \citet{2021ApJ...912..125G}; (36) \citet{1984ApJ...278..754K}; (37) \citet{1980ApJ...239..898F}; (38) \citet{2022MNRAS.511.1574R}; (39) \citet{2015MNRAS.453.1879K}; (40) \citet{2001AA...378..556K}; (41) \citet{2017MNRAS.472.4173R}; (42) \citet{2019RAA....19...88K}; (43) \citet{2006ApJS..167...40E}; (44) \citet{2012MNRAS.425.1394K}; (45) \citet{2013ApJS..204....5K}; (46) \citet{2023ApJS..264...39R}; (47) \citet{2023MNRAS.522..693R}; (48) \citet{1997MNRAS.292..205F}; (49) \citet{2021MNRAS.503.3743W}; (50) \citet{2020MNRAS.497..130T}.} (51) \citet{2021ApJ...912..125G}; (52) \citet{2019MNRAS.483.5291W}; (53) \citet{2020ApJ...894...19R}; (54) \citet{2005MNRAS.357..333B}; (55) \citet{2000MNRAS.315L..41M}; (56) \citet{2021AA...647A.184R}; (57) \citet{1995AA...303L..53D}; 
\end{deluxetable*}
\end{longrotatetable}

Most variables are categorized by their periods of variability (from $100-1500$\,s for pulsating white dwarfs, to minutes to days for spotted white dwarfs), as well as their position in the color-magnitude diagram (e.g., CVs are composed of a composite spectra mixing some fraction of a white dwarf, main-sequence star, and accretion disk, and are considerably over-luminous to an isolated white dwarf). Using the color-magnitude position along with any periods constrained from the survey light curves allows us to suggest likely classifications for the subset of objects in this manuscript; we will present full classification of all 1333 objects in a future publication. 

We devote our attention here to objects where at least one Gaia light curve successfully identifies the period of variability of a candidate white dwarf. We consider Gaia to successfully recover a period if the Gaia-$G$ periodogram has its highest peak matching (to within 0.05 $\mu$Hz) the highest peak of an independent ZTF and/or TESS periodogram, or an independent Gaia bandpass light curve.

Regardless of its significance (we discuss further in Section~\ref{sec:gaiaperiods}), we use the highest-amplitude of the periodogram of the available Gaia, ZTF, and TESS datasets as an initial guess for a linear-least-squares (LLS) fit of a sinusoid to the data using the Python package {\tt lmfit} \citep{2016ascl.soft06014N}. This process returns the best-fit period, amplitude, and phase of a single sinusoid to the light curve, as well as associated uncertainties. All reported periods in the main text of this manuscript are the LLS fits of the Gaia DR3 $G$-band photometry. 

In total we find 105 candidate white dwarfs which have photometric variability periods that can be correctly measured from their Gaia DR3 light curve independently. Of these 105 objects, 72 have data from ZTF and 65 have data from TESS (including 32 with both ZTF and TESS data). Based on our classification (see Section~\ref{sec:wdvariables}), we show where these 105 objects sit in the Gaia color-magnitude diagram in Figure~\ref{fig:cmd}. Because binary systems and objects with cool spots both show sinusiodal variability, we use mass and temperature estimate fits to Gaia photometry by \citet{2021MNRAS.508.3877G} (listed in Table \ref{tab:2}) to distinguish between these objects. We classify candidate white dwarfs with an effective temperature of $>$\,$31{,}000$\,K or a photometrically determined mass $<$\,$0.525$\,\msun\ (which cannot have formed as single-stars within a Hubble time) as binaries; all other sinusoidally variable white dwarfs are classified as rotationally modulated spots. Since not all spotted white dwarfs are strongly magnetic (e.g., \citealt{2017ApJ...835..277H}), the only way to absolutely distinguish between a spot or binary is to use follow-up spectroscopy to search for radial-velocity variability.

Previous spectroscopy exists for just 36 of our 105 targets. Our photometric classifications lacking spectroscopic constraints will suffer some ambiguity. For example, WD\,J202932.51$+$070107.70 is hotter than $31{,}000$\,K and exhibits absorption from ultra-highly excited metals, but is likely variable due to a magnetic spot \citep{2021AA...647A.184R}. Similarly, the $>$\,$100{,}000$\,K white dwarf WD\,J095125.94$+$530930.72 is likely spotted, based on a lack of radial-velocity variation \citep{2019MNRAS.483.5291W}. Still, classifying as binaries the sinusoidal variables with \teff~$>$\,$31{,}000$\,K or ${M}_{\mathrm{WD}}<$\,$0.525$\,\msun\ will be appropriate for the vast majority of our objects.

\section{Classified white dwarf variables} \label{sec:wdvariables}

We attempt to classify the origin of the variability for the 105 candidate variable white dwarfs where we can measure a coherent, sinusoidal variability period from the Gaia light curves that matches a period detected in an independent dataset. Initially we inspect the Gaia DR3 machine learning output classifier ({\tt GaiaDR3Class}), which is available for 62/105 objects. In most cases, however, this classifier simply says {\tt WD}, revealing little about the specific nature of these objects.

We instead classify all objects based on variability period and position in the Gaia color-magnitude diagram, informed by past white dwarf variability studies \citep{2017MNRAS.468.1946H,2018whdw.confE...1R}. All discovery literature on our sources is collected in Table~\ref{tab:2}, and our full classification and variability information is collected in Table~\ref{tab:3}. Table~\ref{tab:summary} provides a summary of the overall sample and our analyzed sub-sample referenced in this manuscript.

\begin{longrotatetable}
\movetabledown=24mm
\begin{deluxetable*}{cHccccccccccccccccccc}
\tablecaption{New variability information determined for the 105 white dwarfs in this study with significant variability measured from their Gaia DR3 light curve.}
\label{tab:3}
\tabletypesize{\tiny}
\tablehead{
\colhead{Class}  & \nocolhead{Gaia Source ID} &\colhead{WD J Name$^{\dagger}$} & \colhead{$P_{\rm G,DR3}$} & \colhead{$\sigma P_{\rm G,DR3}$} & \colhead{$A_{\rm G}$} & \colhead{$\sigma A_{\rm G}$} & \colhead{$t_{\rm min,G}$} & \colhead{$\sigma t_{\rm min}$} & Match & \colhead{$P_{\rm ZTF}$} & \colhead{$\sigma P_{\rm ZTF}$} & \colhead{$A_{\rm ZTF}$} & \colhead{$\sigma A_{\rm ZTF}$} & 
\colhead{$P_{\rm TESS}$} & \colhead{$\sigma P_{\rm TESS}$} & \colhead{$A_{\rm TESS}$} & \colhead{$\sigma A_{\rm TESS}$} & \colhead{Rev.} & \colhead{Std. Rev.} & \colhead{False Alarm} \\ \colhead{(This Work)} & \nocolhead{(DR3)} & & \colhead{[hr]} & \colhead{[hr]} & \colhead{[Rel. Amp.]} & \colhead{[Rel. Amp.]} & \colhead{MJD$_{\rm TDB}$} & \colhead{[d]} & & \colhead{[hr]} & \colhead{[hr]} & \colhead{[Rel. Amp.]} & \colhead{[Rel. Amp.]} &  \colhead{[hr]} & \colhead{[hr]} & \colhead{[Rel. Amp.]} & \colhead{[Rel. Amp.]} & \colhead{[d]} & \colhead{[d]} &\colhead{[\%]}
}
\startdata
Likely Binary&2780903539324180736&WD J003213.13+160434.78&21.771792&0.06451811&0.0053&0.004&56890.44&1.73&ZTF, TESS&21.886956&0.05957178&0.6484&0.3454&19.402039&0.38635639&0.0036&0.0014&52.9&43.2&0.7 \\ Likely Binary&2317319612801004416&WD J003449.86-312952.69&2.314931&0.00036725&0.0357&0.0157&56870.52&0.08&TESS&&&&&2.316131&13.20032778&0.0&0.0015&50.5&40.0&$<$0.1 \\ Likely Binary&522432163970498048&WD J011053.06+604831.03&8.357258&0.00766065&0.007&0.006&56897.32&0.57&TESS&8.373517&0.00404841&0.2803&0.1418&8.362356&0.00191441&0.0643&0.0035&35.3&22.1&$<$0.1 \\ Likely Binary&323342459647016448&WD J012828.99+385436.63&118.136167&0.46444972&0.0155&0.0041&56909.06&2.68&ZTF&119.6655&0.41811806&0.7473&0.1757&122.246833&47.54794444&0.0042&0.0032&61.7&46.9&24.9 \\ Likely Binary&2478893842934875136&WD J013148.02-055900.14&2.055886&0.00033342&0.01&0.0055&56897.97&0.11&TESS&2.056135&0.00019699&0.6549&0.2721&2.025781&0.00804038&0.0096&0.0036&56.7&47.6&19.3 \\ Likely Binary&459237630076876672&WD J022704.24+591502.04&12.736229&0.0122765&0.0073&0.0034&56898.86&&ZTF&12.770364&0.0103218&0.3655&0.177&&&&&38.8&27.5&$<$0.1 \\ Likely Binary&4846154678323893248&WD J032205.10-472106.46&24.768553&0.02279821&0.0393&0.0129&56912.94&0.43&TESS&&&&&23.463694&0.75380139&0.0048&0.0026&30.6&8.3&4.6 \\ Likely Binary&117770030481927936&WD J032803.02+265151.65&8.611367&0.00638363&0.0314&0.0211&56904.43&0.42&ZTF &8.618753&4500.0&0.0&0.5263&&&&&65.6&57.7&28.8 \\ Likely Binary&235184643828418688&WD J032902.63+390340.87&4.036247&0.00198967&0.0242&0.0159&56903.7&0.23&ZTF&4.0431&0.00155644&1.2607&0.6927&&&&&51.2&43.8&5.2 \\ Likely Binary&5090605830056251776&WD J035814.52-213813.74&1.794933&0.00023552&0.0227&0.0118&56914.03&0.09&ZTF&1.794894&0.00033893&0.7248&0.6254&&&&&39.3&29.8&$<$0.1 \\ Likely Binary&4681729620697295360&WD J041033.37-585203.63&2.872539&0.0005393&0.0394&0.0189&56935.79&0.11&TESS&&&&&2.871769&0.00595186&0.002&0.0019&35.9&13.1&15.9 \\ Likely Binary&3204352783173513728&WD J043103.75-041702.63&2.434396&0.00052726&0.0686&0.0404&56912.57&0.11&ZTF&2.434496&0.00043268&2.0772&0.7928&&&&&50.4&46.0&21.9 \\ Likely Binary&3230486971974872192&WD J043832.74+003117.01&26.014614&0.06736786&0.0099&0.0049&56914.98&1.14&ZTF&26.073231&0.11561044&0.1405&0.1963&&&&&43.2&37.5&1.1 \\ Likely Binary&2974680502235811712&WD J050433.01-213007.93&15.028869&0.02485904&0.0143&0.0107&56918.23&0.87&ZTF&14.956056&0.01491123&1.5712&0.9047&&&&&36.6&23.7&4.4 \\ Likely Binary&3341182435406679424&WD J053632.07+123128.65&3.801928&0.00249101&0.0102&0.0126&56912.42&0.46&ZTF, TESS&3.799225&0.00054618&0.8572&0.2873&3.840864&0.03100203&0.002&0.0023&47.9&47.7&$<$0.1 \\ Likely Binary&2941800362921100928&WD J061234.26-191111.75&2.224975&0.00058326&0.0143&0.0134&56919.44&0.16&ZTF&2.225262&0.00038728&1.3633&0.9595&&&&&36.7&27.1&1 \\ Likely Binary&5549322119820590208&WD J061815.27-512305.52&9.674728&0.00529894&0.0112&0.004&56932.46&0.31&TESS&&&&&9.645339&0.00446996&0.0097&0.001&35.0&11.1&28.6 \\ Likely Binary&3125897813875563904&WD J064316.01+013012.78&6.036606&0.00212576&0.0224&0.0097&56918.12&0.24&ZTF, TESS&6.041864&0.00277571&0.5138&0.3596&6.038822&0.00072595&0.0091&0.0015&60.1&49.8&22.3 \\ Likely Binary&5575611644702244992&WD J064753.57-403756.16&7.011806&0.0106999&0.0142&0.0172&56996.71&0.67&Gaia BP&&&&&&&&&32.5&10.0&$<$0.1 \\ Likely Binary&5286227228018008576&WD J064802.27-634629.90&3.775269&0.00143243&0.0166&0.0146&56931.78&0.2&TESS&&&&&3.765814&0.00170344&0.005&0.0015&34.9&10.6&$<$0.1 \\ Likely Binary&1099005032089977984&WD J070647.52+613350.31&8.965739&0.00733039&0.0044&0.0021&56904.96&0.45&ZTF, TESS&8.956211&0.00547992&0.1324&0.0848&8.963867&0.00529426&0.0002&0.0001&44.7&33.5&7 \\ Likely Binary&989059981050056320&WD J072146.74+564408.41&2.715027&0.00056963&0.0218&0.012&56907.27&0.14&ZTF&2.714582&0.0013891&1.2807&1.4401&&&&&51.8&44.8&3.6 \\ Likely Binary&1085903148454238208&WD J073121.39+595737.01&5.315978&0.00316432&0.0104&0.0106&56906.67&0.33&ZTF, TESS&5.309414&0.00956094&0.1098&0.3165&5.313603&19.30418611&0.0&0.0013&46.9&37.6&33.7 \\ Likely Binary&5615892360564187008&WD J074316.04-224656.86&8.310972&0.00983921&0.019&0.0187&56959.09&0.64&ZTF&8.279997&0.00263236&4.3029&1.709&&&&&38.1&19.1&$<$0.1 \\ Likely Binary&5216924803961432192&WD J090240.78-742331.37&5.449119&0.0018044&0.0101&0.0045&56893.28&0.21&TESS&&&&&5.416497&0.02826039&0.0045&0.0026&33.9&11.8&9 \\ Likely Binary&3842377248804727168&WD J091748.20+001041.72&29.146361&0.08106744&0.0119&0.0097&56962.01&1.71&ZTF, TESS&29.178917&0.03566194&1.0295&0.4407&42.345667&4.53270833&0.009&0.0049&49.6&44.6&0.5 \\ Likely Binary&5437786045395659520&WD J093611.12-340059.16&20.793195&0.0359738&0.0067&0.0043&56965.09&&TESS&&&&&16.100917&0.39214361&0.033&0.0097&36.4&20.0&$<$0.1 \\ Likely Binary&3834895969825628800&WD J100645.76+003204.48&2.536604&0.00066172&0.0242&0.0185&56959.73&0.13&ZTF, TESS&2.535475&0.00043236&0.5741&0.3772&2.536101&0.00034043&0.0023&0.0013&57.0&50.2&17.6 \\ Likely Binary&5456112705203412352&WD J105228.94-295308.20&8.229394&7527.777778&0.0&0.0001&56874.03&432563.6&TESS&&&&&8.361094&0.12081919&0.0189&0.0056&50.6&36.1&27.6 \\ Likely Binary&5390573840730111872&WD J110318.00-401544.50&29.047667&0.0949295&0.0272&0.0174&56881.84&1.83&TESS&&&&&29.240944&2.59028361&0.0008&0.0004&47.7&31.5&5.9 \\ Likely Binary&5383832838018956032&WD J112355.21-400614.63&16.972742&0.01447909&0.0288&0.0131&56887.85&0.45&TESS&&&&&16.983225&135.3446944&0.0&0.0&45.4&30.1&12.8* \\ Likely Binary&5378958286359002624&WD J113832.79-435817.73&13.311775&0.03923664&0.004&0.008&56891.91&1.75&Gaia BP, Gaia RP&&&&&&&&&39.7&25.8&$<$0.1 \\ Likely Binary&6076202559948311040&WD J121622.61-545907.30&2.576151&0.00050299&0.0664&0.035&56894.91&0.11&Gaia BP, Gaia RP&&&&&&&&&34.1&17.6&0.1 \\ Likely Binary&1726297924930902400&WD J123955.40+834707.51&0.956676&0.00035594&0.0388&0.0896&56898.18&0.19&ZTF&0.956609&0.00014681&4.9308&5.1684&&&&&47.2&22.8&7.4 \\ Likely Binary&3509435507186636288&WD J125005.11-203632.29&21.147906&0.02900806&0.0345&0.0262&56896.21&0.91&TESS&21.082511&0.03706747&0.9748&0.9013&21.486694&1.88802694&0.0084&0.0065&62.8&50.7&12 \\ Likely Binary&3721943209023827456&WD J135523.92+085645.42&2.744749&0.0006954&0.0132&0.0078&56901.83&0.17&ZTF, TESS&2.744139&0.00024016&0.7035&0.2419&2.780175&0.02776044&0.0084&0.0055&53.3&48.4&0.1 \\ Likely Binary&1671880139535534208&WD J140330.89+671255.16&3.443786&0.00139008&0.0168&0.0122&56897.06&0.24&ZTF&3.4503&0.00066051&0.6072&0.3348&&&&&36.0&11.9&3.1 \\ Likely Binary&1286055427676135808&WD J144007.60+315413.02&23.245372&0.04633281&0.0135&0.0083&56913.96&1.1&ZTF, TESS&23.265067&0.01444292&0.7476&0.2314&23.246072&0.01754755&0.0206&0.0051&33.5&14.0&9.3 \\ Likely Binary&5773385817714683776&WD J145623.30-793654.99&4.819789&0.0018614&0.0221&0.0159&56898.7&0.22&Gaia BP&&&&&&&&&32.5&10.2&$<$0.1 \\ Likely Binary&5873013322220969984&WD J145817.91-651024.94&3.677164&0.00275866&0.004&0.0062&56900.11&0.5&Gaia BP&&&&&&&&&31.9&16.9&$<$0.1 \\ Likely Binary&1728880879608190592&WD J153028.82+874848.48&21.206697&0.02246878&0.0186&0.007&56904.12&0.54&TESS&&&&&21.425442&0.01431244&0.0013&0.0003&32.5&8.7&$<$0.1 \\ Likely Binary&1272870835853944704&WD J154012.07+290828.96&2.377268&0.00026954&0.0467&0.0179&56914.8&0.07&ZTF&2.377583&0.000173&1.8722&0.5728&&&&&30.5&16.8&$<$0.1 \\ Likely Binary&2121629147470776704&WD J181824.35+465721.46&3.730808&0.00084071&0.0454&0.0209&56930.63&0.14&ZTF&3.735769&0.00101204&0.374&0.2643&&&&&37.8&13.4&0.9 \\ Likely Binary&2095490942878080256&WD J183243.45+354518.42&1.416905&0.00019062&0.0107&0.0079&56927.91&0.08&ZTF&1.416682&5.93e-05&0.4607&0.1244&&&&&32.7&10.5&0.4 \\ Likely Binary&2144580907240525568&WD J184506.31+505412.15&5.742764&0.00158156&0.0332&0.0131&56931.08&0.16&ZTF, TESS&5.75205&0.00177661&0.2939&0.1531&5.739953&0.00302524&0.0211&0.0079&32.6&7.3&$<$0.1 \\ Likely Binary&2126940892444693760&WD J191931.09+443243.88&6.209089&0.00248988&0.0163&0.0074&56928.17&0.22&ZTF, TESS&6.21685&0.0016627&0.2733&0.1146&6.234617&0.00393408&0.0038&0.0016&33.9&9.6&0.3 \\ Likely Binary&1807309147099442432&WD J195952.94+143534.17&7.188675&0.0166399&0.0029&0.0059&56992.27&1.04&ZTF&7.191394&0.00177492&0.8598&0.3362&&&&&42.2&30.9&0.4 \\ Likely Binary&1809460380271929344&WD J200937.02+165219.12&2.498782&0.00086119&0.0176&0.0127&56958.77&0.15&ZTF&2.500491&0.00017069&2.4724&0.5959&2.547105&0.02237542&0.0085&0.0055&40.0&29.4&$<$0.1 \\ Likely Binary&2061416695278519808&WD J202034.57+392212.12&5.324653&0.00283764&0.0192&0.0106&56990.84&0.28&TESS&&&&&5.320283&0.00187972&0.024&0.0072&31.7&9.4&2.4 \\ Likely Binary&6679158828046846336&WD J202354.90-433513.67&15.2327&0.02066798&0.0214&0.0114&56914.23&0.69&TESS&&&&&16.172639&0.25935103&0.0278&0.0049&51.3&41.7&$<$0.1 \\ Likely Binary&1863757043284422784&WD J202841.38+325837.35&19.422681&0.02613492&0.0656&0.0348&56963.42&0.66&ZTF, TESS&19.616389&0.02075202&2.0971&1.0358&21.614186&0.15849081&0.5815&0.0392&32.7&12.6&$<$0.1 \\ Likely Binary&6910676050839449728&WD J210110.17-052751.14&12.874175&0.01180351&0.0077&0.0031&56957.41&0.45&Gaia BP, Gaia RP&12.673861&90555.55556&0.0&0.0066&&&&&51.5&43.6&0.5 \\ Likely Binary&2177297391130977792&WD J210840.53+554502.63&2.781769&0.00053615&0.0281&0.0141&56873.09&0.09&ZTF&2.782606&0.0006235&0.7013&0.5455&&&&&37.2&16.3&0.3 \\ Likely Binary&1867851869401000960&WD J211557.58+364255.40&6.753503&0.00451239&0.0125&0.0088&56963.93&0.36&ZTF&6.748281&0.00153208&1.049&0.3375&&&&&32.7&13.4&0.3 \\ Likely Binary&1950573244454108160&WD J213119.93+345246.84&10.373233&0.01306096&0.0105&0.0079&56988.5&0.61&ZTF&10.412589&0.00437928&1.075&0.3873&&&&&37.1&22.2&$<$0.1 \\ Likely Binary&2667317799127474944&WD J213958.29-074126.17&6.250269&0.00316512&0.0244&0.0108&56983.85&0.25&ZTF&6.261194&0.00157262&1.4742&0.6431&&&&&53.2&46.4&$<$0.1 \\ Likely Binary&2000785539622175104&WD J221545.15+503759.46&5.414369&0.00237553&0.0224&0.0144&56882.25&0.25&ZTF&5.431119&0.00513291&1.6915&1.2771&&&&&35.7&13.0&12.2 \\ Likely Binary&2833849800205759360&WD J224539.30+201633.35&3.905189&0.00094606&0.0096&0.0051&56867.64&0.11&ZTF&3.905517&0.00084593&0.1878&0.0921&&&&&60.1&45.2&6.6 \\ Likely Binary&2608280729159380992&WD J224653.73-094834.48&19.826454&0.01028305&0.0452&0.0036&56958.88&&TESS&19.837572&0.16538997&1.1215&0.5601&18.505856&0.11626889&0.036&0.0024&57.0&50.1&32.1 \\ Likely Binary&6609580078678720512&WD J225206.39-274012.56&5.132572&0.0023864&0.0258&0.0159&56954.48&0.31&ZTF&5.133864&0.00253317&0.6989&0.6005&&&&&58.8&51.9&4.6 \\ Likely Binary&2411529617359368960&WD J230047.06-125054.98&16.714308&0.02454316&0.0129&0.007&56959.68&0.93&TESS&16.589397&0.01605333&0.4336&0.2738&16.554344&0.01973398&0.0091&0.0027&62.4&52.3&27.1 \\ Likely Binary&2210180554096762624&WD J232411.16+653053.05&7.271892&0.00379897&0.0196&0.009&56895.47&0.33&ZTF, TESS&7.303556&0.00417417&0.491&0.3447&7.314125&0.00476089&0.0057&0.0011&31.8&8.9&$<$0.1 \\ Likely Binary&6523373973310305664&WD J233923.79-485350.19&2.529616&0.00074594&0.0091&0.0078&56912.48&0.14&TESS&&&&&2.521869&0.01775451&0.002&0.0021&36.4&27.2&$<$0.1 \\ Likely Eclipsing Binary&414813439007636352&WD J004502.35+503408.46&6.883219&0.00192382&0.0772&0.0294&56927.53&0.16&ZTF&6.869319&0.00253022&1.1385&0.4951&&&&&41.8&31.2&40.7 \\ Likely Eclipsing Binary&6086116512688802176&WD J130905.43-470852.39&3.320381&0.00107816&0.0589&0.0493&56896.94&0.22&Gaia RP&&&&&&&&&52.6&38.5&27 \\ Likely Eclipsing Binary&6008982469163902464&WD J154008.28-392917.61&1.31965&0.00016547&0.0486&0.0276&56903.93&0.06&TESS&&&&&1.323371&0.00525961&0.01&0.0055&46.0&39.4&28.5 \\ Likely Eclipsing Binary&2056368017063855360&WD J202811.64+350525.08&10.405553&0.00659212&0.0212&0.0081&56963.46&0.34&ZTF&10.396467&27611.11111&0.0&0.3689&&&&&32.7&13.7&1.8 \\ Known Eclipsing Binary&380560941677424768&WD J003352.64+385529.70&4.861628&0.00576532&0.0082&0.0111&56895.07&0.84&ZTF&4.862839&0.0011144&0.9914&0.4022&&&&&52.6&39.9&$<$0.1 \\ Known Eclipsing Binary&578539413395848704&WD J085746.18+034255.35&1.561434&0.00047608&0.0307&0.0246&56956.65&0.14&ZTF&1.561681&0.0001148&2.6183&1.2482&&&&&49.5&44.8&$<$0.1 \\ Known Eclipsing Binary&1446230735421615872&WD J132518.18+233808.02&4.678903&0.0018771&0.0127&0.0074&56886.94&0.25&ZTF&4.674961&0.0018441&0.6449&0.5285&&&&&48.2&37.5&0.2 \\ Known Binary&1112171030998592256&WD J071709.79+740040.53&2.472226&0.00034023&0.0078&0.0032&56902.62&0.08&ZTF, TESS&2.474291&0.00035304&0.1626&0.0756&2.473012&0.00029588&0.0006&0.0003&33.1&14.5&21 \\ Known Binary&5660008787157868928&WD J095017.97-251124.79&7.644542&0.00560631&0.0145&0.0063&56963.73&0.34&TESS&&&&&7.290883&0.11245208&0.0027&0.0011&51.1&38.6&$<$0.1 \\ Known Binary&1223988576107848832&WD J153938.12+270605.84&5.72125&0.00534064&0.0032&0.0033&56914.25&0.55&TESS&5.721197&0.00086724&0.598&0.2038&5.727303&0.00233915&0.0028&0.0017&33.5&17.1&$<$0.1 \\ Likely CV&6180386608127234432&WD J130017.86-325805.23&0.603127&9.41e-05&0.1924&0.2752&57021.15&0.08&Gaia BP, Gaia RP&&&&&&&&&54.9&44.6&20.2 \\ Known CV&5099482805904892288&WD J031413.25-223542.92&1.349474&0.00016573&0.0242&0.0178&56912.08&0.07&ZTF, TESS&1.350296&0.00015006&2.167&1.0267&1.360255&0.00542343&0.0099&0.0053&50.1&37.3&29.8 \\ Known CV&5645504991845223936&WD J083843.35-282701.14&16.109283&0.02880356&0.0893&0.0609&56960.29&0.99&Gaia BP, Gaia RP&&&&&&&&&31.6&16.7&$<$0.1 \\ Likely DAV&5709854047291834496&WD J082924.77-163337.25&0.099723&1.77e-06&0.0079&0.0086&56959.48&0.01&ZTF&0.099723&6.9e-07&1.195&0.7069&&&&&48.6&34.4&2.5 \\ Likely DAV&3459790225027896448&WD J120650.89-380549.23&0.072198&3.6e-07&0.0079&0.004&56893.04&0.0&TESS&&&&&0.072209&1.86e-05&0.0018&0.0011&47.4&37.3&8.1 \\ Known DAV&2026653131220381824&WD J195227.88+250929.18&0.07111&7.6e-07&0.0036&0.0031&56960.12&0.01&ZTF, TESS&0.071112&3.3e-07&0.2168&0.1663&0.071111&3e-07&0.0009&0.0006&33.9&17.1&3.7* \\ Known DAV&6410501923531836928&WD J215823.88-585353.81&0.086112&5.8e-07&0.0079&0.0037&56907.42&0.0&TESS&&&&&0.086111&8.4e-07&0.0003&0.0003&37.9&22.1&28.9** \\ Likely Spot&313206676129937408&WD J011157.12+313632.67&9.814925&0.01994233&0.0043&0.0043&56897.32&1.42&ZTF&9.821594&0.00612669&0.3733&0.2035&&&&&57.6&49.9&12.3 \\ Likely Spot&5139880551029408768&WD J013832.02-195446.58&3.941892&0.00119064&0.0051&0.003&56891.56&0.16&ZTF&3.943017&0.00196275&0.3572&0.3089&3.942033&0.00057022&0.0099&0.003&43.4&35.1&$<$0.1 \\ Likely Spot&105601048101988992&WD J020322.94+255851.90&25.724439&0.04739078&0.0346&0.0185&56893.56&0.86&TESS&&&&&31.755639&5.15051111&0.0047&0.0044&65.7&55.8&10.2 \\ Likely Spot&4618860233015905408&WD J024235.52-820636.32&3.342828&0.00118396&0.0068&0.0049&56897.31&0.21&TESS&&&&&3.339775&0.00235079&0.0008&0.0006&33.8&12.5&1.2 \\ Likely Spot&187576000803144704&WD J052026.35+381240.87&6.785561&0.00458165&0.0084&0.0051&56908.17&0.38&ZTF, TESS&6.795878&0.00127562&0.6634&0.1692&6.830475&0.04033464&0.0067&0.001&61.4&53.6&24.6 \\ Likely Spot&2951380789991336192&WD J062555.04-141442.31&4.459511&0.00126425&0.0193&0.0075&56920.07&0.15&ZTF, TESS&4.463656&0.00153086&0.7396&0.4556&4.734681&0.06312842&0.0027&0.0014&47.8&37.3&$<$0.1 \\ Likely Spot&3161618477052648192&WD J072758.87+101157.08&0.491355&2.25e-05&0.0063&0.0055&56918.41&0.03&ZTF&0.491394&1.51e-05&0.4658&0.2585&&&&&64.3&54.5&25.1 \\ Likely Spot&1030289712881895296&WD J083531.17+533230.91&2.163019&0.00042838&0.0135&0.009&56909.2&0.12&ZTF&2.161886&0.00042694&1.6366&1.0268&&&&&49.1&40.6&33.7 \\ Likely Spot&5620763437599189376&WD J090212.89-394553.32&3.560872&0.00073476&0.0081&0.0028&56964.15&0.1&TESS&&&&&3.642272&0.05924833&0.0018&0.0016&31.6&10.6&0.6 \\ Likely Spot&3598424931753893888&WD J115033.35-063618.07&0.245574&2.62e-06&0.0118&0.0039&56876.94&0.0&ZTF, TESS&0.245557&6.9e-06&0.2916&0.3757&0.245566&1.33e-05&0.0011&0.0008&56.5&48.7&9.8 \\ Likely Spot&1465214177336901376&WD J124740.10+303007.41&225.656194&1.46535889&0.0216&0.0057&56878.2&3.13&ZTF&228.333778&2.97987222&0.508&0.2783&1545.405833&377777.7778&0.0474&11.0918&51.9&39.8&2.2 \\ Likely Spot&6279014454702485632&WD J144410.08-214338.33&2.120309&0.0011789&0.0037&0.0061&56902.83&0.33&ZTF&2.120568&0.00021151&0.6368&0.328&&&&&65.6&50.7&35.5 \\ Likely Spot&1639907647071426176&WD J154801.37+623611.57&4.736575&0.00205755&0.0224&0.0148&56875.26&0.2&ZTF&4.736792&0.00128098&0.7656&0.4439&&&&&34.2&11.1&0.1 \\ Likely Spot&4334641562477923712&WD J165335.21-100116.33&79.865139&0.29106056&0.0143&0.0038&56922.1&1.61&ZTF&79.478111&0.579565&0.5255&0.578&&&&&43.2&39.6&$<$0.1 \\ Likely Spot&2043899619620928000&WD J185924.85+330157.53&1.634021&3.47e-05&0.354&0.006&56926.42&&ZTF&1.634515&0.00050539&0.5311&0.4539&&&&&35.2&18.1&40.7* \\ Likely Spot&2216598403463898112&WD J215432.26+634154.11&0.934589&6.78e-05&0.0244&0.015&56892.59&0.04&ZTF&0.934499&5.65e-05&0.6797&0.4345&&&&&33.8&10.6&0.6 \\ Likely Spot&6613841171565998848&WD J221452.56-314333.98&7.392242&0.00398858&0.0142&0.0071&56951.62&0.35&TESS&&&&&6.902656&0.13466103&0.0063&0.003&49.5&43.3&0.2 \\ Known Spot&2375576682347401216&WD J003512.88-122508.84&0.605106&6.89e-06&0.078&0.0054&56879.66&&ZTF, TESS&0.60499&1.93e-05&0.4396&0.2066&0.604392&26.48977778&0.0&0.0&65.5&51.2&25.3 \\ Known Spot&4613612951211823104&WD J031715.85-853225.56&0.201399&1.92e-06&0.0362&0.013&56897.19&0.01&TESS&&&&&0.201389&0.00105979&0.0&0.0003&32.5&8.8&$<$0.1 \\ Known Spot&551153263105246208&WD J041246.85+754942.26&0.344425&9.77e-06&0.0066&0.0036&56937.29&0.12&ZTF, TESS&0.344443&1.04e-05&0.2687&0.1765&0.344439&5.52e-06&0.0013&0.0006&33.1&14.6&4.2 \\ Known Spot&504240778439455872&WD J053432.93+770757.40&0.723311&7.13e-05&0.0076&0.0085&56900.79&0.06&ZTF, TESS&0.723223&16.23804444&0.0&0.0064&0.72334&0.00451125&0.0&0.0004&35.6&13.5&5.5* \\ Known Spot&1020641700210987520&WD J095125.94+530930.72&82.886&1.00760111&0.0132&0.0119&56918.71&6.77&ZTF, TESS&82.097083&0.5039775&0.7059&0.509&83.268694&0.14100106&0.0038&0.0007&36.4&25.6&$<$0.1 \\ Known Spot&3682469122383597056&WD J125230.93-023417.72&0.088057&0.00133891&0.0&0.0055&56890.23&7.57&TESS&&&&&0.088&5.57e-05&0.0023&0.003&54.4&45.2&0.1 \\ Known Spot&4235280071072332672&WD J195629.23-010232.67&34.898417&0.07376844&0.016&0.0051&56917.58&1.05&TESS&&&&&34.644361&0.40300639&0.0047&0.0003&48.0&41.6&$<$0.1 \\ Known Spot&1748833236082347136&WD J202932.51+070107.70&6.957214&0.00222029&0.0159&0.0045&56959.12&0.18&ZTF&6.971394&0.00149846&1.11&0.3097&7.140053&0.06677989&0.0173&0.0044&44.0&36.0&13.8 \\
\enddata
\tablerefs{$^{\ast}$Gaia BP ($G_{\mathrm{BP}}$) band used in analysis instead of Gaia G band. $^{\ast\ast}$Gaia RP ($G_{\mathrm{RP}}$) band used in analysis instead of Gaia G band. Note that ``Rev.'' (revisit) is the average time between two Gaia detections of a source, and ``Std. Rev.'' is the standard deviation of this revisit time.}
\end{deluxetable*}
\end{longrotatetable}

\subsection{Non-Eclipsing White Dwarf Binaries} \label{subsec:binaries}

Binary systems are often distinguishable from isolated white dwarfs because they are an unresolved pair of stars, making them over-luminous to single star cooling tracks, such as those shown in Figure~\ref{fig:cmd}. For this reason, over-luminous objects with sinusiodal variability that have an effective temperature exceeding $31{,}000$\,K as determined by the hydrogen-rich-atmosphere fits from \citet{2021MNRAS.508.3877G} or a photometrically determined mass $<0.525$\,\msun\ are categorized as binary systems. Objects without mass determinations from \citet{2021MNRAS.508.3877G} are likely over-luminous and also classified as binaries. Typical values for orbital periods of binary systems with a white dwarf paired with an M dwarf have been studied by \citet{2011A&A...536A..43N}, where SDSS data showed that post common envelope binaries containing a white dwarf and main sequence star have orbital periods ranging from 1.9\,hr to 103.2\,hr, with an average of approximately 10.3\,hr. 

We find 72 likely variable binaries in our sample, including seven eclipsing binaries (Section~\ref{subsec:ebs}), with orbital periods ranging from 0.957\,hr to 29.366\,hr, with a mean of approximately 8.411\,hr and a standard deviation of approximately 7.417\,hr, consistent with the \citet{2011A&A...536A..43N} distribution.

\begin{figure}[t]
\centering{\includegraphics[width=0.995\columnwidth]{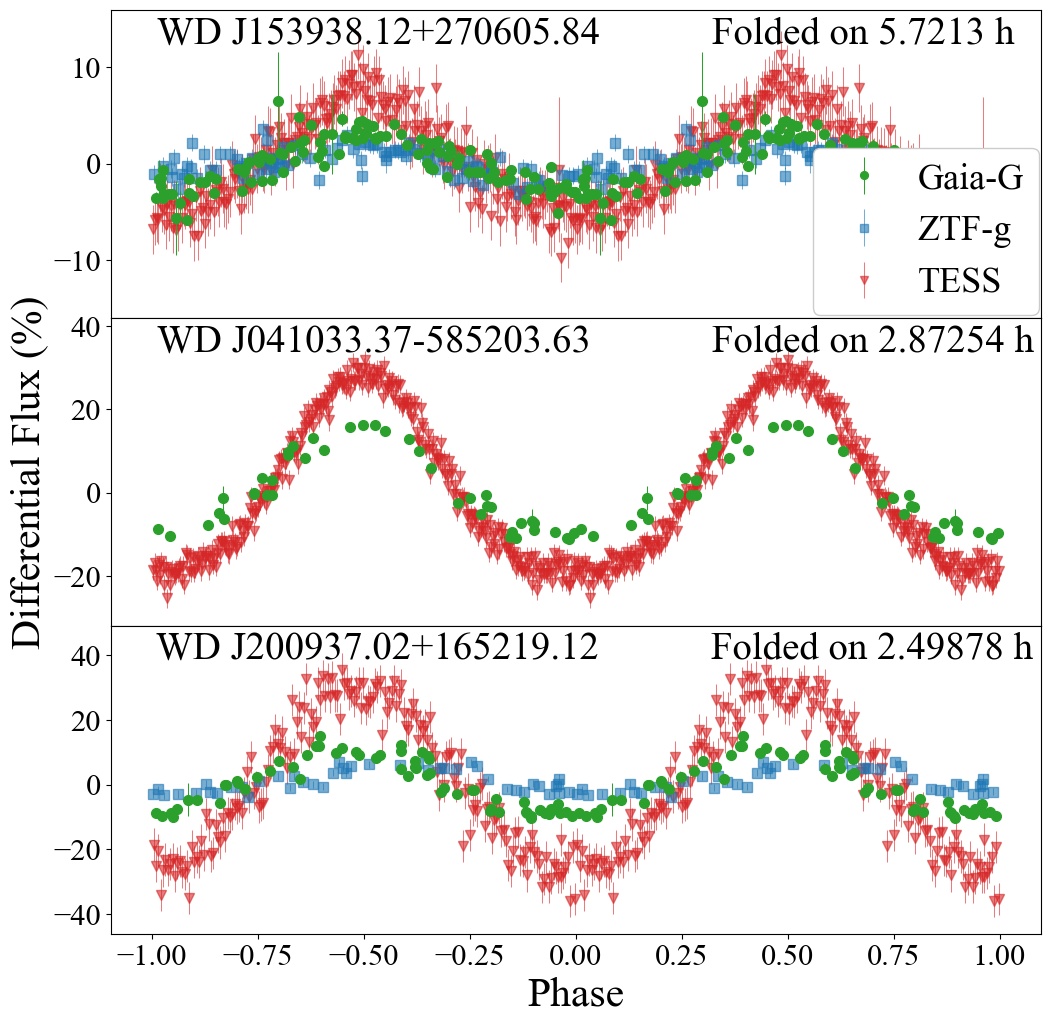}}
\caption{Phase-folded light curves of three representative variable white dwarfs from binarity, revealing orbital periods of 5.725\,hr, 2.873\,hr, and 2.500\,hr, respectively. All folded light curves (ZTF $g$ in blue, Gaia $G$ in green, and TESS photometry in red) are phased to the same ephemeris. ZTF $g$ and TESS data are grouped into 151 bins for each light curve. The larger amplitudes at redder wavelengths in these systems indicates they are all likely reflection-effect binaries of a cool main-sequence star reprocessing flux from a nearby hot white dwarf.  \label{fig:binaries}}
\end{figure}

Three representative examples of variable binaries are shown in Figure~\ref{fig:binaries}. The first object (WD\,J153938.12$+$270605.84, $G$ = 16.6\,mag) is a known DA+dM system with an orbital period of $5.72120\pm0.00087$\,hr measured from the Palomar Transient Factory \citep[PTF,][]{2016MNRAS.461.2747K}. We measure from Gaia DR3 photometry only $P_{\mathrm{orb}}=5.7213\pm0.0053$\,hr, agreeing with the PTF measurement to within our estimated uncertainty. The second object (WD\,J041033.37$-$585203.63, $G$ = 17.6\,mag) is a newly discovered binary system with a sinusiodal pattern in both the Gaia $G$ and TESS bands with a Gaia $G$ band period of $2.87254\pm0.00054$\,hr. The higher amplitude from the redder TESS data is indicative of flux reprocessing from a reflection-effect binary, though no spectroscopy exists for this target. Finally, WD\,J200937.02$+$165219.12 ($G$ = 16.8\,mag) is also likely a new reflection-effect binary with an orbital period of $2.49878\pm0.00086$\,hr. 

The shortest-period, likely detached new binary in our sample is WD\,J123955.40$+$834707.51 ($G$ = 20.0\,mag), which shows significant variability in ZTF at $0.95661\pm0.00015$\,hr that is also seen in the Gaia $G$-band light curve at an amplitude exceeding 40\%. No spectroscopy exists of this target, which has a position in the Gaia CMD consistent with a 7020\,K, 0.27\,\msun\ white dwarf, suggesting it might be a new WD+WD binary containing at least one ELM white dwarf (e.g., \citealt{2022ApJ...933...94B}).

\subsection{Periodicities in Known Cataclysmic Variables}\label{sec:CVs}

There are also likely some binaries that are not detached in our sample. CVs are a class of accreting white dwarf binaries, and are often eruptive variables that show large brightening events from dwarf novae and other outbursts \citep{2020AJ....159..198S, 2021AJ....162...94S}. We have not intended to search for new CVs among the 1333 high-confidence white dwarfs, which has been the focus of previous work on Gaia light curves using eruptions and outbursts (e.g.,  \citealt{2019AA...623A.110G}). However, we have identified coherent periods in three objects classified as CVs in the literature. 

WD\,J083843.35$-$282701.14 ($G$ = 17.6\,mag) is a known CV \citep{2017ApJ...838..124H, 2017MNRAS.471.2902R} for which we find a new potential period of variability of $16.109\pm0.029$\,hr. This period differs slightly from the approximately 15-hr period of flux variability interpreted as the beat period between the orbital and spin periods \citep{2017MNRAS.471.2902R}, as well as the $14.7\pm1.2$\,hr period of X-ray modulations observed by \citet{2017ApJ...838..124H}. WD\,J031413.25$-$223542.92 ($G$ = 18.2\,mag) was studied by \citet{1980MNRAS.191P..43W} and \citet{1979ApJ...234L.113B}. The Gaia DR3 data show a significant periodicity at $1.34947\pm0.00017$\,hr, which is in close agreement with both studies.

Finally, WD\,J130017.86$-$325805.23 ($G$ = 20.3\,mag), is a CV candidate that has undergone at least two measured outbursts detected in CRTS \citep{10.1093/mnras/stu639}. We lack TESS or ZTF coverage for this faint southern object, but all three Gaia band-passes have their highest periodogram peak at 2171.36\,s (36.2\,min). The amplitude of the signal exceeds 80\% in all Gaia band-passes. Such large-amplitude periodic variations are not commonly observed as spurious signals in Gaia DR3 at such short periods \citep{2023AA...674A..25H}, though we cannot independently rule out an instrumental origin for this 36.2-min variability. We strongly encourage follow-up of this faint southern target. 

\subsection{Eclipsing White Dwarf Binaries} \label{subsec:ebs}

\begin{figure}
\centering{\includegraphics[width=0.995\columnwidth]{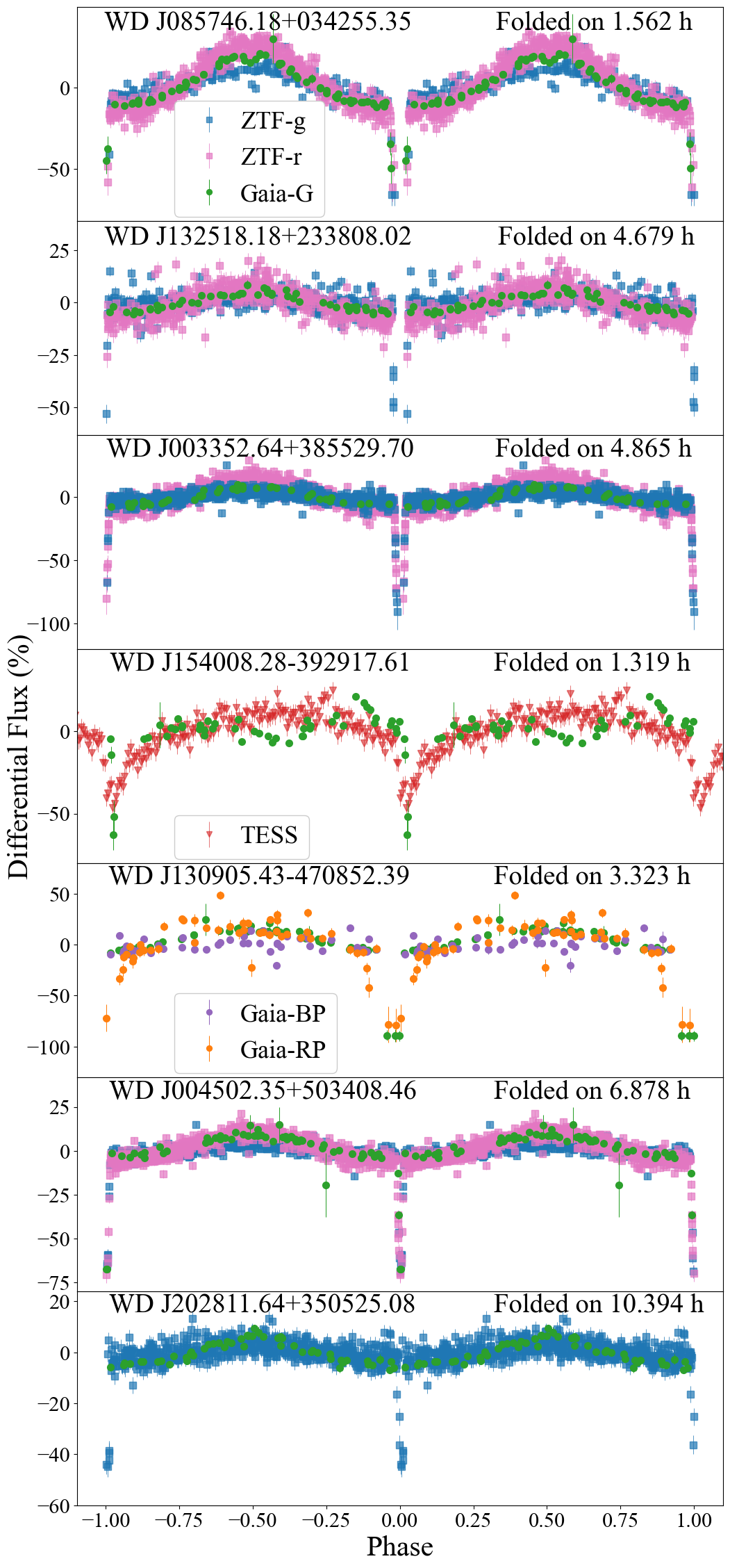}}
\caption{Following the same convention as Figure~\ref{fig:binaries}, we show phase-folded light curves of seven eclipsing binaries detected from their Gaia DR3 light curves. In this figure, ZTF and Gaia data are not binned, but TESS data are grouped into 251 bins. The first three systems are previously published eclipsing binaries containing a white dwarf \citep{2010arXiv1009.3048D,2015MNRAS.449.2194P,2022MNRAS.516..720K}. Four systems are newly discovered eclipsing binaries containing a white dwarf, including a new 1.3-hr eclipsing CV (WD\,J154008.28$-$392917.61, further detailed in Section~\ref{subsubsec:FollowUp} and  Figure~\ref{fig:follow_up}). With only sparse Gaia DR3 data, we do not claim to confirm binarity in WD\,J154008.28$-$392917.61, but include it as an interesting candidate.
\label{fig:eclipsing}}
\end{figure}

Eclipsing binaries often display near-total flux dropouts in randomly sampled light curves  \citep{2013MNRAS.429..256P}, as the fainter companion passes in front of the brighter member. These eclipses should occur at exactly the same phase in a folded light curve, manifesting as stable, repeated photometric variability. The seven known and likely eclipsing systems are shown in Figure~\ref{fig:eclipsing}.

The top panels in Figure~\ref{fig:eclipsing} detail three previously known eclipsing white dwarf binaries, in order of orbital period. WD\,J085746.18$+$034255.35 ($G$ = 18.3\,mag) is a known eclipsing WD+dM with a measured orbital period of 1.562317\,hr by the Catalina Sky Survey (CSS, \citealt{2015MNRAS.449.2194P}), which is consistent with our measurement from Gaia DR3 photometry of $1.562316(17)$\,hr. This object shows both ZTF and Gaia dropouts, notable since the Gaia DR3 light curves have a reasonably aggressive outlier rejection \citep{2023A&A...674A..13E}. WD\,J132518.18$+$233808.02 ($G$ = 18.2\,mag) is also a known eclipsing DA+dM binary with an orbital period of 4.679\,hr measured by CSS \citep{2010arXiv1009.3048D,2015MNRAS.449.2194P} consistent with our measurement from Gaia DR3 photometry of $4.67901(39)$\,hr. The lack of in-eclipse data from Gaia is likely related to outlier rejection of the Gaia photometry\footnote{\href{https://gea.esac.esa.int/archive/documentation/GDR3/Data_processing/chap_cu5pho/cu5pho_sec_photProc/cu5pho_ssec_photCal.html}{Gaia DR3 Documentation Section 5.1.4: Photometric Calibration}}. WD\,J003352.64$+$385529.70 ($G$ = 18.3\,mag) was identified by \citet{2022MNRAS.516..720K} to have a strong 4.86-hr signal in ZTF, and has all the hallmarks of a WD+dM eclipsing binary, especially featuring ZTF dropouts in both the $g$- and $r$-bands when folded on a period of $4.86465(31)$\,hr. These dropouts align in phase space and show a decrease in amplitude of nearly 100\%.

Our analysis identifies four new candidate eclipsing binaries. Most compact is WD\,J154008.28$-$392917.61, showing multiple low Gaia points when folded on the TESS-determined period of $1.319(13)$\,hr. We detail follow-up observations of this source in Section~\ref{subsubsec:FollowUp} and Figure~\ref{fig:follow_up}.

WD\,J130905.43$-$470852.39 ($G$ = 18.3\,mag) may be an eclipsing binary, having only Gaia light curves to analyze. When folded on a period of $3.32265(44)$\,hr, the source shows  dropouts in both the $G$ and $G_{\mathrm{RP}}$ bands with small uncertainties leading us to believe that these are genuine eclipses. Photometric fits to the Gaia CMD suggest this system has an extremely low-mass ($<$0.16\,\msun), 11040 K white dwarf with a high probability of being a white dwarf ($P_{\rm WD} = 0.91$, \citealt{2021MNRAS.508.3877G}). With only sparse Gaia DR3 data and possibly long transit durations, we cannot confirm this system as eclipsing yet, but encourage further follow-up. 

We are far more confident in the fidelity of the final two systems shown in Figure~\ref{fig:eclipsing}. WD\,J004502.35$+$503408.46 ($G$ = 17.6\,mag) is over-luminous to the isolated white dwarf position in the Gaia CMD (suggesting binarity) and features ZTF and Gaia dropouts suggesting an orbital period of $6.87794(72)$\,hr. WD\,J202811.64$+$350525.08 ($G$ = 17.9\,mag), is also over-luminous and features ZTF dropouts when folded on a period of $10.3937(17)$\,hr. There is also a strong Gaia-$G$ sinusiodal pattern that aligns with the ZTF data with a minimum in the time of transit.

\subsubsection{Follow-Up of a New Eclipsing CV} \label{subsubsec:FollowUp}

We obtained follow-up time-series photometry and spectroscopy for our most compact new eclipsing system (WD\,J154008.28$-$392917.61), located in the southern hemisphere ($G$ = 17.3\,mag).

We obtained high-speed time-series photometry using the ULTRACAM instrument \citep{2007MNRAS.378..825D} mounted on the 3.6-meter New Technology Telescope (NTT) telescope at the La Silla Observatory in Chile. This camera obtains simultaneous three-channel photometry with frame-transfer detectors; we used high-throughput Super-SDSS $u_s g_s r_s$ filters \citep{2021MNRAS.507..350D}. We observed WD\,J154008.28$-$392917.61 over two consecutive nights, analysing here only the SDSS-$g_s$ data. We collected 9.0-s exposures on UT date 2023~March~16 with an average seeing of 0.8\,arcsec. We collected 4.5-s exposures on UT date 2023~Mar~17 with an average seeing of 0.7\,arcsec. All ULTRACAM data were bias-subtracted and flat-field corrected using the HiPERCAM pipeline\footnote{\href{https://github.com/HiPERCAM/hipercam}{https://github.com/HiPERCAM/hipercam}}.

We also obtained follow-up spectroscopy of this system using the Goodman spectrograph \citep{2004SPIE.5492..331C} mounted on the 4.1-meter SOAR Telescope in Chile. We collected eleven $10$\,min spectra using the 400-line grating with a 1-arcsec slit and the M1 grating setup on UT date 2023~May~24, yielding a resolution of roughly 4.4\,\AA. We extracted the data using the {\tt pypeit} data reduction pipeline \citep{2020JOSS....5.2308P}, and flux calibrated using the standard star EG~274.

\begin{figure}[t]
\centering{\includegraphics[width=0.995\columnwidth]{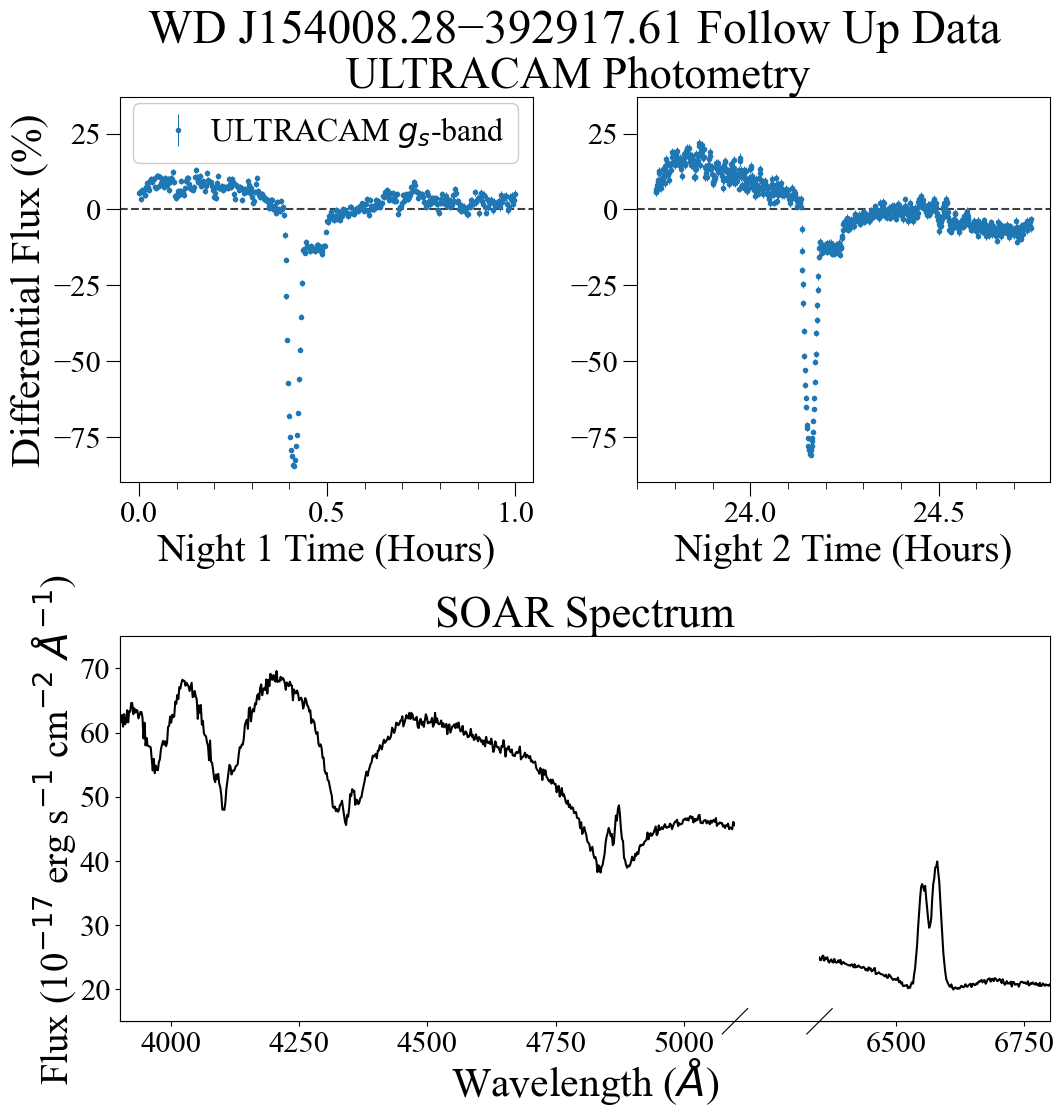}}
\caption {Follow up data for WD\,J154008.28$-$392917.61 reveals deep primary eclipses in the two top panels. Follow-up spectroscopy (note the broken x-axis) confirms this is a 68.365-min CV, likely descending from a previously detached white dwarf + brown dwarf post-common-envelope binary (e.g., \citealt{2007MNRAS.381..827L}).} 
\label{fig:follow_up}
\end{figure} 

We captured two primary eclipses with ULTRACAM, shown in the top two panels of Figure~\ref{fig:follow_up}. Fitting the mid-time of the eclipses with a Gaussian bounded by when the flux decreases below $-$20\%, we find that the primary eclipses occur at BJD$_{\rm TDB}$ = 2460019.8887179(17) and 2460020.8778468(11), separated by 18 cycles of our TESS-determined main periodicity of $1.319(13)$\,hr. There is considerable out-of-transit variability, including an extended feature from a bright spot on the white dwarf, seen in many eclipsing CVs (e.g., \citealt{2011MNRAS.415.2025S,2019MNRAS.486.5535M}).

Our coadded spectrum in the bottom panel of Figure~\ref{fig:follow_up} confirms this is an accreting white dwarf binary. Broad Balmer-line absorption from the white dwarf is visible, along with strong emission features from accretion, especially dominant at H$\alpha$. Our spectrum is not high-enough resolution to attempt a radial-velocity fit. We do not observe Zeeman splitting at the Balmer lines, though our limits are weak given the strong emission in the line cores.

A purported orbital period of 68.365\,min is well below the CV period minimum of roughly 80\,min, after which convective donor stars expand on further mass loss \citep{1984ApJS...54..443P,2011ApJS..194...28K,2016ApJ...833...83K,2019MNRAS.486.5535M}. However, the full photometric phase coverage from TESS shown in Figure~\ref{fig:eclipsing} reveals this is the correct period of recurrent variability. This makes WD\,J154008.28$-$392917.61 among the shortest-period CVs known, similar to the 66.61-min CV SDSS\,J150722.30$+$523039.8 \citep{2007MNRAS.381..827L}. These ultracompact CVs are likely descended from a previously detached white dwarf + brown dwarf binary. It is also possible the system descended from a low-metallicity, significantly evolved donor \citep{2011MNRAS.414L..85U}.

\subsection{Magnetic Spotted White Dwarfs} \label{subsec:spots}

Spotted white dwarfs are difficult to distinguish from reflection-effect binaries like those discussed in Section~\ref{subsec:binaries}. When spectroscopy is not available, we classify all objects with apparently mono-periodic variability and photometrically determined \citep{2021MNRAS.508.3877G} masses $>$0.525\,\msun\ and temperatures $<$$31{,}000$\,K as isolated white dwarfs with rotational modulation from magnetic spots.

\begin{figure}[t]
\centering{\includegraphics[width=0.995\columnwidth]{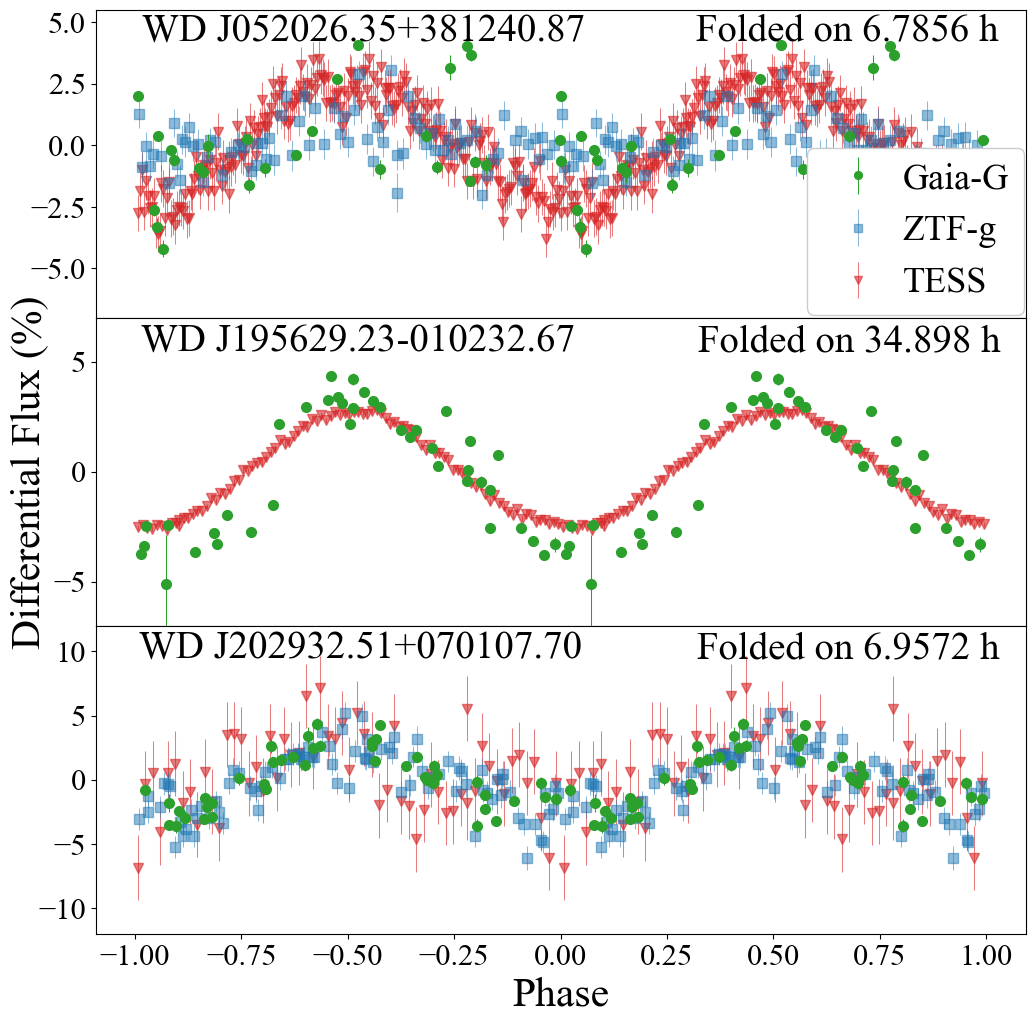}}
\caption{Following the same convention as Figure~\ref{fig:binaries}, we show phase-folded light curves of three magnetic spotted white dwarfs. ZTF-$g$ and TESS data are grouped into 151 phase bins, whereas the Gaia data are not binned. \label{fig:spots}}
\end{figure}

The period of the spot-induced modulations corresponds to the rotational period at the star's surface (e.g., \citealt{2013ApJ...773...47B,2015ApJ...814L..31K,2018AJ....156..119H}). We recover the periods of eight previously known spotted white dwarfs from the Gaia DR3 light curves independently, and identify 17 new likely spotted white dwarfs. The majority of the newly identified spotted white dwarfs (12/17) have effective temperatures between $6500-10{,}000$\,K, placing them inside the range of known DAHe strongly magnetic white dwarfs with Balmer-line emission \citep{1985ApJ...289..732G,2020ApJ...894...19R,2020MNRAS.499.2564G, 2021MNRAS.503.3743W, 2023MNRAS.522..693R, 2023MNRAS.521.4976M}, motivating future time-resolved spectroscopic follow-up. Our sample contains several known DAHe (see Table~\ref{tab:2}). WD\,J083531.17$+$533230.91 ($G$ = 18.2\,mag) is the only other spectroscopically confirmed strongly magnetic white dwarf in our sample, and appears to have a 2.16-hr period. 

Examples of three spotted white dwarfs are shown in Figure~\ref{fig:spots}. The top object WD\,J052026.35$+$381240.87 ($G$ = 15.4) is a spectroscopically confirmed white dwarf from LAMOST claimed to be a DBA \citep{2019RAA....19...88K} but is likely a DAH; Gaia DR3 photometry shows it is likely rotating at $6.7856\pm0.0046$\,hr. The second object (WD\,J195629.23$-$010232.67, $G$ = 13.6\,mag) has been known to be spotted (the white dwarf is highly magnetic with a complex field structure, shown in \citealt{2000MNRAS.315L..41M}). We observe clear sinusoidal variability in the $G$ data with a period of $34.898\pm0.074$\,hr, in close agreement with the literature value from \citet{2005MNRAS.357..333B}. The final object is WD\,J202932.51$+$070107.70 ($G$ = 16.6\,mag), a known DO white dwarf \citep{1995AA...303L..53D}. This white dwarf was observed to have a ZTF $g$ photometric variability period of $6.97882\pm0.00012$\,hr from \citet{2021AA...647A.184R}, which is close to our LLS fit to the Gaia-$G$ data of $6.9572\pm0.0022$\,hr. It is likely that the ZTF or Gaia data has selected an incorrect alias due to gaps in the data.

\begin{figure}[t]
\centering{\includegraphics[width=0.995\columnwidth]{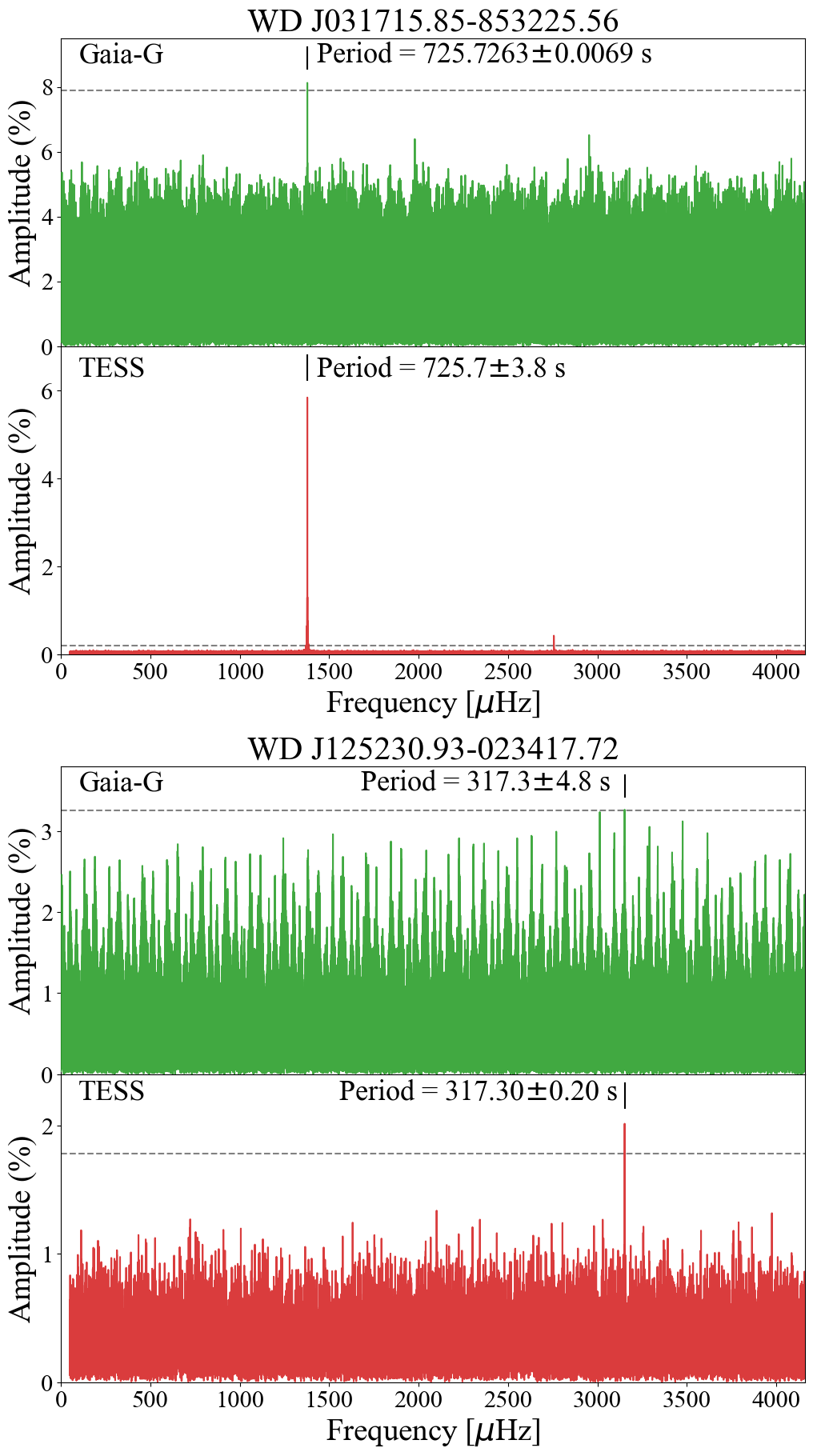}}
\caption{Lomb-Scargle periodograms showing two rapidly rotating spotted white dwarfs recovered in sparsely sampled Gaia DR3 light curves. At top is WD\,J031715.85$-$853225.56
  \citep{1997MNRAS.292..205F}; the 725\,s variability period has a $<$0.1\% false-alarm probability in the Gaia $G$ light curve. At bottom is WD\,J125230.93$-$023417.72, which has a 0.1\% false-alarm probability, with the highest peak in the periodogram at exactly the same period, within the uncertainties, as the known variability discovered by \citet{2020ApJ...894...19R}.} \label{fig:spotshortper}
\end{figure}

Figure~\ref{fig:spotshortper} demonstrates Gaia DR3 light curves are able to recover the variability in two of the shortest-period spotted white dwarfs known. The top panel details  WD\,J031715.85$-$853225.56 ($G$ = 14.8\,mag), a well-studied, rapidly rotating, high-field magnetic white dwarf \citep{1997MNRAS.292..205F}. The Gaia $G$ peak occurs at the same period as the TESS data, to within the uncertainties of the LLS fits. The light curve from Gaia is sparsely sampled, with visits separated by $32.5\pm8.8$\,d, where 8.8\,d is the standard deviation of the time between visits. False-alarm probability lines for periodograms are estimated by holding the time sampling the same but randomly shuffling the fluxes with replacement 10{,}000 times, in an attempt to quantify the probability that the amplitude of a peak can be generated by pure noise (e.g., \citealt{2014MNRAS.438.3086G,2018ApJS..236...16V}). For  WD\,J031715.85$-$853225.56 at the top of Figure~\ref{fig:spotshortper}, none of our bootstrapped periodograms have a peak as high as the observed Gaia $G$ periodogram, yielding a false-alarm probability of $<$0.1\%.

In the bottom panel of Figure~\ref{fig:spotshortper} we show Gaia and TESS peaks measured for the strongly magnetic, emission-line (DAHe) white dwarf WD\,J125230.93$-$023417.72 ($G$ = 17.5\,mag, \citealt{2020ApJ...894...19R}). While not statistically significant, the highest peak in the Gaia periodogram occurs at the known dominant period of 317.278\,s (which is the first harmonic the 634.56-s rotation period, see \citealt{2023MNRAS.525.1097F}). We bootstrap a false-alarm probability of 0.1\% for this peak.

Aside from these two previously known spotted white dwarfs, we recover the literature periods for six other known spots. Two are known emission-line DAs: the 36.3-min DAHe system 
WD\,J003512.88$-$122508.84 \citep{2023MNRAS.522..693R} and the 2.29-hr, non-magnetic DAe white dwarf WD\,J041246.84$+$754942.26 \citep{2020MNRAS.497..130T, 2023MNRAS.524.4996E}. We recover two cooler white dwarfs with spots: the 0.72-hr WD\,J053432.93$+$770757.40 \citep{2021ApJ...912..125G} and the 1.44-d WD\,J195629.23$-$010232.67 shown in Figure~\ref{fig:spots}. We also recover two much hotter spotted stars discussed at the end of Section~\ref{sec:analysis}, the 6.98-hr WD\,J202932.51$+$070107.70 \citep{2021AA...647A.184R} and the 3.45-d WD\,J095125.94$+$530930.72
 \citep{2019MNRAS.483.5291W}.

The shortest-period new spotted white dwarf in our sample is WD\,J115033.35$-$063618.07 ($G$ = 16.4\,mag). This spotted white dwarf features a highest peak in the $G$ light curve matching the TESS-significant peak at $884.0679\pm0.0094$\,s. Spectropolarimetry puts a 3$\sigma$ upper limit on the magnetic field of LP 673-41 of $<$6\,kG \citep{2012MNRAS.425.1394K}, raising questions about the origin of the spot on this 8600\,K white dwarf. The longest period spot appears on WD\,J124740.10$+$303007.41 ($G$ = 17.9\,mag), with a highest peak in the periodogram at 9.487\,d in both its Gaia and ZTF light curves.

\subsection{Pulsating White Dwarfs} \label{subsec:zzs}

\begin{figure*}[t]
\centering{\includegraphics[width=0.995\textwidth]{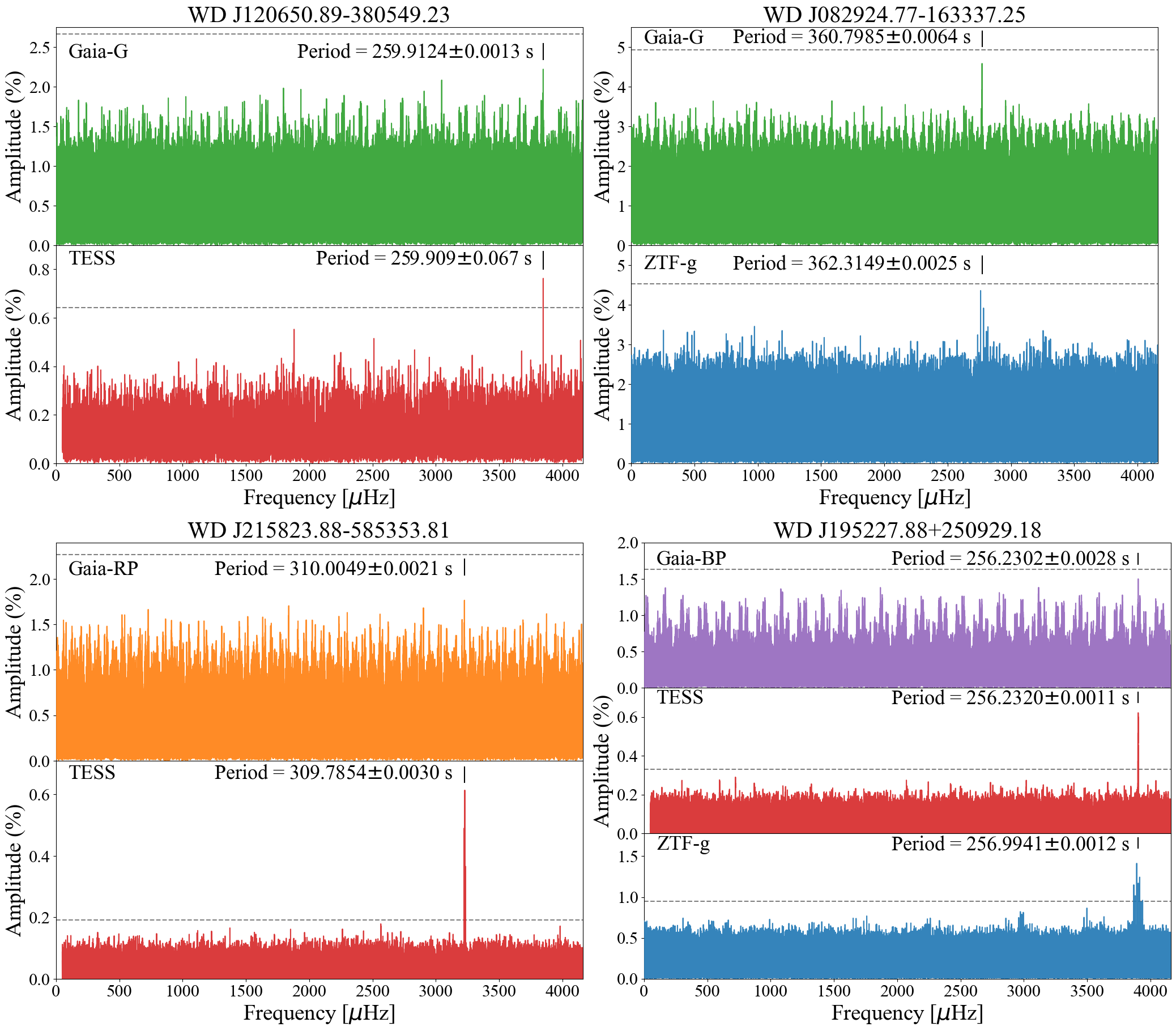}}
\setlength{\belowcaptionskip}{12pt}
\caption{Panels of periodograms for the four objects we classify as Gaia-detected pulsating white dwarfs, all with periods $<$400\,s. The photometric bands in these figures are $G$, $G_{\mathrm{BP}}$, $G_{\mathrm{RP}}$, $g$, $r$, $T$ (Gaia G, BP, and RP bands, ZTF g and r bands, and the TESS photometry band, respectively). A bootstrapped 99.9\% confidence lines is shown as a horizontal dashed line in gray (see description in Section~\ref{subsec:spots}). The false-alarm probabilities of all four Gaia peaks are all between 2.5\% and 28.9\% (see Table~\ref{tab:3}).} \label{fig:zzs}
\end{figure*}

Pulsating white dwarfs are variable due to non-radial pulsations, with periods typically ranging from $100-1500$\,s, amplitudes up to several percent, and usually multiple, non-harmonically related periods present at the same time \citep{2008PASP..120.1043F,2008ARAA..46..157W}. DA white dwarfs are unstable to pulsations in a narrow temperature range between roughly $13{,}500-10{,}500$\,K \citep{2017ApJS..232...23H}. These hydrogen-atmosphere pulsators are commonly known as ZZ Cetis and DAVs.

From Gaia DR3 photometry alone we are able to independently identify the periods for two new candidate pulsating white dwarfs and confirm the periods of two known DAVs. All objects lie within the empirical DAV instability strip in the Gaia color-magnitude diagram (Figure~\ref{fig:cmd}), and their Lomb-Scargle periodograms are shown in Figure~\ref{fig:zzs}. As with our spotted white dwarfs in Figure~\ref{fig:spots}, we bootstrap false-alarm probabilities to assess the significance of a peak in the periodogram and quantify the likelihood a peak cannot arise from noise alone. The significance lines in Figure~\ref{fig:zzs} are the 99.9\% confidence line ($<$0.1\% false alarm); false-alarm probabilities for all objects in this manuscript are listed in Table~\ref{tab:3}. While none of our pulsating white dwarfs have formally significant peaks, the chance that the highest peak in the periodogram exactly matches significant peaks in an independent ZTF or TESS dataset is exceptionally low; our periodograms all have step sizes of 0.01\,$\mu$Hz and run beyond 4000\,$\mu$Hz, giving hundreds of thousands of trial frequencies. In three of the four cases in Figure~\ref{fig:zzs}, the false-alarm probability is $<$10\%.   

We find WD\,J120650.89 $-$380549.23 ($G$ = 15.7\,mag) is a new pulsating WD with a pulsation period at $259.9127\pm0.0013$\,s in $G$ with a 8.1\% false-alarm probability that is also significantly detected in TESS. WD\,J215823.88-585353.81 ($G$ = 15.8\,mag) is a known DAV with multiple detected modes from rotational splittings revealed by TESS at 310.27\,s, 309.79\,s, and 309.31\,s \citep{2022MNRAS.511.1574R}. The highest peak in the $G_{\mathrm{RP}}$ periodogram is consistent with this pulsational variability, at a period of $310.0049\pm0.0021$\,s, though this peak has a marginal 28.9\% false-alarm probability. WD\,J082924.77$-$163337.25 ($G$ = 17.6\,mag) was identified as a variable white dwarf in \citet{2021ApJ...912..125G}. We detect variability consistent with pulsations in both the Gaia-$G$ and ZTF-$g$ light curves at $359.0045\pm0.0025$\,s, with just a 2.5\% false-alarm probability. Finally, WD\,J195227.88$+$250929.18 ($G$ = 15.1\,mag), is a known DAV with a dominant pulsation period of 256.2\,s \citep{1984ApJ...278..754K,1980ApJ...239..898F}. Our analysis agrees closely with a peak in $G_{\mathrm{BP}}$ at $255.999\pm0.001$\,s and a 3.7\% false-alarm probability of matching peaks in ZTF and TESS.

The four white dwarfs with likely pulsations detected in their Gaia light curves all have periods $<$400\,s; longer-period modes are likely unstable over long baselines \citep{2017ApJS..232...23H}, suggesting longer-period modes will be hard to detect from Gaia data independently. However, pulsating white dwarfs appear to be the most numerous white dwarfs flagged as variable in Gaia DR3. There is an overdensity of points in Figure~\ref{fig:cmd} near the DAV instability strip relative to the 200-pc sample of white dwarfs in the background. In fact, of the 883 candidate white dwarfs that are flagged as photometrically variable in Gaia DR3 with $P_{\mathrm WD}>0.98$ and masses $>$0.45\,\msun\ from \citet{2021MNRAS.508.3877G}, nearly half (422) have effective temperatures near the DAV instability strip boundaries ($13{,}500 > $\teff$ > 10{,}500$\,K). Literature searches of this sample reveal at least 50 are already-known DAVs, and at least 120 that can be photometrically confirmed with TESS, ZTF or ASAS-SN data; this analysis will be presented in a future publication.

\section{Gaia-Detected Periods} \label{sec:gaiaperiods}

We use Gaia DR3 epoch photometry, which includes photometric time series data in the $G$-, $G_{\mathrm{BP}}$-, and $G_{\mathrm{RP}}$- bands \citep{2023A&A...674A..13E}. Each time-series data point is a so-called field-of-view transit, after an object has passed through nine consecutive CCD observations, each separated by 4.85\,s \citep{2018AA...620A.197R}. As the Gaia spacecraft spins, objects will then be re-observed at a cadence separated by 1\,hr~46\,min \citep{2017MNRAS.472.3230R}. The time sampling after this revisit is considerably longer, and different for each source, set by the Gaia scanning law (e.g., \citealt{2021MNRAS.501.2954B}). 

\begin{figure}[t]
\centering{\includegraphics[width=0.995\columnwidth]{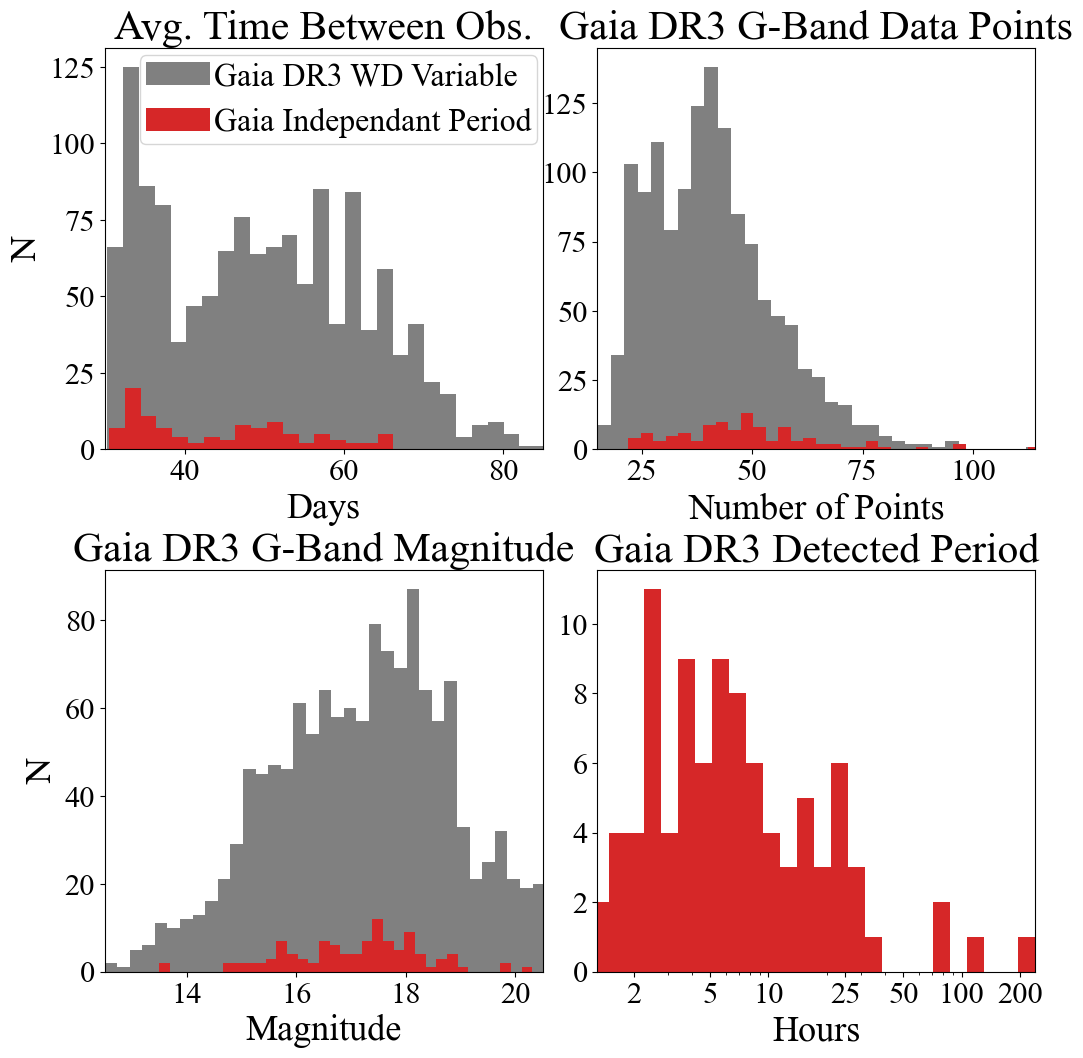}} 
\caption{Top left histogram shows the average time between Gaia DR3 observations (ignoring the 1\,hr~46\,min revisit) for the full sample of 1333 candidate white dwarfs flagged as variable in gray, inset in red by the 105 candidate white dwarfs in this manuscript where their variability period can be determined by Gaia independently. The top right panel shows the number of points in each $G$ light curve, and the bottom left panel shows the $G$ magnitude distribution of both samples. The bottom right panel shows the distribution of the 105 periods for objects detected by Gaia independently (note that the x-axis is log scale). Although we are slightly more sensitive to objects with more frequent visits, our sample of 105 sources covers most of this parameter space. \label{fig:histogram}}
\end{figure}

We analyze the revisit cadence of these observations by creating a histogram of the average time between each observation, shown in Figure~\ref{fig:histogram}. We ignore the revisit timescale of 1\,hr~46\,min (set by the 6-hr spin period of Gaia, as discussed in \citealt{2012Ap&SS.341...31D}) by excluding time between observations of less than 0.3 days, and find an average time of $49.7\pm12.5$\,d between observations for the full sample of 1333 objects (the uncertainty is set by the standard deviation of these revisit times). Of the 105 objects where we can measure variability from the Gaia light curves independently, the average revisit time is $44.2\pm10.3$\,d. Our subset of 105 objects has slightly less time between observations, only a few more Gaia-G data points (with an average number of points of $49\pm17$ compared to $41\pm14$ for the full sample), and is slightly brighter (with an average magnitude of $G= 17.0\pm1.2$\,mag compared to $G = 17.2\pm1.6$\,mag for the full sample). The distributions in Figure~\ref{fig:histogram} show we are not heavily biased towards objects with frequent revisits set by the Gaia scanning law. The average and standard deviation in the revisit times between FoV transits for each object is included in Table~\ref{tab:3}. Among the 72 of 105 objects with ZTF data, the average object has 393 ZTF visits, which is completely consistent with the typical sky coverage of the average fields in ZTF DR12.

We also analyze short-period variability in the Gaia DR3 light curves, which is measured as the object moves from one CCD to the next every 4.85\,s (e.g., \citealt{2018AA...620A.197R}). Included in DR3 are per-CCD least-squares fits listed as ``Frequency'' for objects flagged as {\tt vari\_short\_timescale}, for select objects with candidate variability periods shorter than 0.5\,d. The inverse of the {\tt vari\_short\_timescale} frequency is listed in Table~\ref{tab:2} as $P_{\rm G,S}$, and we check each period generated by our LLS fit ($P_{\rm G,DR3}$ in Table~\ref{tab:3}) against $P_{\rm G,S}$. In our final 105 object sample, 65 objects are flagged in Gaia DR3 as short-timescale sources, and 55 of these 65 (85\%) have $P_{\rm G,S}$ that matches $P_{\rm G,DR3}$ to within 3\% (after which the agreement is never better than 22\%).

We compare the Gaia DR3 short-timescale analysis against our LLS fits by object type. We find that of our 72 binary objects, 48 of 53 (91\%) with {\tt vari\_short\_timescale} data have $P_{\rm G,S}$ that matches our determinations of $P_{\rm G,DR3}$, and of the spotted white dwarfs, 5 of 7 with short-timescale analysis match. For all variable types, the Gaia DR3 short-timescale analysis returns 55/60 (92\%) matches for objects with {\tt vari\_short\_timescale} between 0.278 and 2.78 hours, and routinely does poorly for periods longer than 3\,hr.

We also attempt to assess the reliability of the  Gaia variability identifier \citep{2023A&A...674A..13E}. Gaia uses a machine learning algorithm trained on non-variable objects and then tested against training sets (built from various published catalogues of variable stars) to confirm how effectively this algorithm can identify variable objects. Objects are categorized as variable when there is a large value for the ratio of the difference in magnitude from the average for that object to the interquartile range of the magnitude. These objects are then passed through one of two paths in the machine learning process to be sorted into a specific Gaia ``class,'' discussed in depth in \citet{2023A&A...674A..14R}. This Gaia class is listed in Table~\ref{tab:2}.

We find that this ``class'' is not generally useful, as most are simply ``WD.'' There is one case where this class correctly identifies a known CV (WD\,J130017.86$-$325805.23). The Gaia ``class'' also marks two objects as ``S'' or stars with short-timescale variability: WD\,J083843.35$-$282701.14 (a known CV with $P_{\rm G,S}$ that matches $P_{\rm G,DR3}$ at $16.109\pm0.029$\,hr) and WD\,J233923.79$-$485350.19 (a likely new binary system with $P_{\rm G,S}$ that matches $P_{\rm G,DR3}$ at $2.52962\pm0.00075$\,hr). The new eclipsing CV WD\,J154008.28$-$392917.61 also has a Gaia ``class'' of CV. 

Finally, we note that Gaia epoch photometry may contain spurious periodicities due to scan-angle-dependent signals, especially in marginally resolved, fixed-orientation pairs of stars separated by $<$0.5 arcsec \citep{2023AA...674A..25H}. Although previous work has only explored signals up to 10 days, and most signals determined here are below that, we investigate the likelihood that any of our detections are spurious. 

All 1333 white dwarf candidates in our sample have information in the {\tt vari\_spurious\_signals} table from \citet{2023AA...674A..25H}, including important metrics that can reveal close binarity, the Spearman rank correlation coefficient with corrected flux excess factor ($r_{\rm exf,G}$) and the Image Parameter Determination model ($r_{\rm ipd,G}$). Only three white dwarfs have both $r_{\rm exf,G}$ and $r_{\rm ipd,G}$ values $>$0.8 (WD\,J020426.03$+$571026.62, WD\,J031047.57$-$100153.85, and WD\,J100643.74$-$221757.86), none of which are included in our subset of 105 white dwarfs with variability measured independently from Gaia. In our subset of 105 objects, only one has $r_{\rm ipd,G}>0.6$ and that is WD\,J003512.88$-$122508.84, which has confirmed variability from TESS \citep{2023MNRAS.522..693R}. We conclude that contamination from spurious Gaia scanning-law-dependent periods is unlikely in our sample of white dwarfs.

\section{Discussion \& Conclusions} \label{sec:summary}

We analyze 105 white dwarf candidates for which Gaia is capable of detecting accurate periods from DR3 photometry. These periods of variability range from 9.5\,d to as short as 256.2\,s (roughly 4\,min), including seven objects with periods shorter than 1000\,s. We obtained this result by comparing Gaia multi-band photometry to data from ZTF and TESS to identify significant periods within the Gaia data. Analysis of the Gaia photometry resulted in the discovery of new pulsating, spotted and binary white dwarfs. This includes a new 68.4-min eclipsing cataclysmic variable, which has a period well below the CV period minimum of roughly 80\,min \citep{2011ApJS..194...28K} and is one of the shortest-period CVs known.

Our overall analysis demonstrates that Gaia epoch photometry is precise enough to detect short-term, low-amplitude periods even when observations are sparsely sampled. The absolute median amplitude of the photometric variability we confirm from Gaia independently is $1.4\%$, with an average period of roughly 12\,hr. This is notable compared to the distribution of time between observations, which is on average roughly 44\,d. We find that overall, the 105 variables that are significant in Gaia are slightly brighter and sampled more often than the full sample of 1333 objects, but the distributions of magnitudes or number of points are not significantly biased in either parameter space, as shown in Figure~\ref{fig:histogram}.

The likelihood that the highest peak in two independent periodograms match is small, given that there are more than $10^{5}$ frequency elements in each periodogram. This is the basis for which we claim Gaia has detected a periodicity. Still, we have bootstrapped estimates that a peak in the periodogram can be caused by noise alone and find that 33/105 objects have significant peaks in their Gaia light curves, with a $<$0.1\% false-alarm probability. Nearly half the sample, 50/105, have peaks with a false alarm $<$1\%. Only three have $>$35\%, but all 105 have Gaia peaks with a $<$41\% false-alarm probability (Table~\ref{tab:3} details this value for all objects in our sample). Our work provides an empirically determined set of cutoffs for which variability in a Gaia DR3 light curve is likely to be genuine, and it also represents a useful value for future analysis of sparsely sampled, high-precision photometry from large surveys like the Vera C. Rubin Observatory and its Legacy Survey of Space and Time (LSST; \citealt{2019ApJ...873..111I}), as well as the Nancy Grace Roman Space Telescope \citep{2019arXiv190205569A}.

Previous spectroscopy exists for only 36 of the 105 objects discussed in this paper, so follow up is encouraged for most of these sources. There are many objects of interest that would expand sample sizes of rare object types. Some notable examples include WD\,J123955.40$+$834707.51 ($G$ = 20.0\,mag), which may be a new WD+WD binary containing an ELM white dwarf, and WD\,J130017.86$-$325805.23 ($G$ = 20.3\,mag) which is a candidate CV with recorded outbursts \citep{10.1093/mnras/stu639} but shows a dominant photometric period of 36.2\,min closer to the orbital period of AM CVn binaries (see Section~\ref{subsec:binaries}). There are also multiple candidate eclipsing binaries discussed in Section~\ref{subsec:ebs}. Finally, one of the shortest-period new spotted white dwarfs we discovered here (WD\,J115033.35$-$063618.07) appears to have an 884.73-s spin period but a $<$6\,kG magnetic field, raising questions about the origin of its fast rotation (see Section~\ref{subsec:spots}).

As Gaia continues to scan the sky, the methods discussed in this paper provide a foundation for analyzing short-period variables within even longer-baseline data. Verifying that Gaia data independently can be used to detect short periods will allow future work on sources with no additional spectroscopic or photometric information. Increasing the numbers of rare variable stellar remnants, especially those in short-period binaries or with magnetic spots, will help constrain their space density and formation channels.

\section{acknowledgements} \label{sec:ack} 

The authors thank the anonymous referee for a very useful report, as well as fruitful discussions with Stu Littlefair. Support for this work was in part provided by NASA TESS Cycle 2 Grant 80NSSC20K0592 and Cycle 4 grant 80NSSC22K0737, as well as the National Science Foundation under grant No. NSF PHY-1748958, as portions of this project were completed at the KITP Program ``White Dwarfs as Probes of the Evolution of Planets, Stars, the Milky Way and the Expanding Universe.''

The authors acknowledge the European Southern Observatory program 110.23UU.001, as well as the Transiting Exoplanet Survey Satellite (TESS), the Zwicky Transiting Facility (ZTF), the All-Sky Automated Survey for Supernovae (ASAS-SN), and the Asteroid Terrestrial-impact Last Alert System (ATLAS) for the public release of all-sky photometry data for many of these objects. This work has also made use of data from the European Space Agency (ESA) mission
{\it Gaia} (\url{https://www.cosmos.esa.int/gaia}), processed by the {\it Gaia}
Data Processing and Analysis Consortium (DPAC,
\url{https://www.cosmos.esa.int/web/gaia/dpac/consortium}). Funding for the DPAC
has been provided by national institutions, in particular the institutions
participating in the {\it Gaia} Multilateral Agreement.

This research is also based on observations obtained at the Southern Astrophysical Research (SOAR) telescope, which is a joint project of the Minist\'{e}rio da Ci\^{e}ncia, Tecnologia e Inova\c{c}\~{o}es (MCTI/LNA) do Brasil, the US National Science Foundation’s NOIRLab, the University of North Carolina at Chapel Hill (UNC), and Michigan State University (MSU).

\vspace{5mm}
\facilities{The Gaia Collaboration, The Zwicky Transient Facility, The Transiting Exoplanet Survey Satellite, NTT, SOAR}

\software{astropy \citep{2013A&A...558A..33A,2018AJ....156..123A}, numpy \citep{2020Natur.585..357H}, lightkurve \citep{2018ascl.soft12013L}, pandas \citep{2022zndo...3509134R}, astroquery \citep{2019AJ....157...98G}, matplotlib \citep{2007CSE.....9...90H}, lmfit \citep{2016ascl.soft06014N}, TOPCAT \citep{2005ASPC..347...29T}}

\bibliography{dr3_variable_wds}{}
\bibliographystyle{aasjournal}

\end{document}